\documentclass[letterpaper,11pt]{article}
\usepackage[T1]{fontenc}
\usepackage[margin=1in]{geometry}
\usepackage{setspace}
\usepackage{amsmath,amssymb}
\DeclareUnicodeCharacter{2011}{\nobreakdash-}
\DeclareUnicodeCharacter{2013}{--}
\DeclareUnicodeCharacter{2014}{---}
\DeclareUnicodeCharacter{201C}{``}
\DeclareUnicodeCharacter{201D}{''}
\usepackage{siunitx}
\usepackage{graphicx}
\usepackage{subcaption}
\usepackage{booktabs}
\usepackage{overpic}
\usepackage{multirow}
\usepackage{xcolor}
\usepackage{authblk}
\usepackage[hidelinks]{hyperref}
\usepackage[backend=biber,style=chem-acs,sorting=none,articletitle=true,doi=true]{biblatex}
\title{Multi-Fidelity Machine Learning for Predicting Startup
Stress Growth and Strain Hardening in Wormlike Micellar Solutions}
\author[1]{Saghar Zolfaghari\textsuperscript{\dag}}
\author[2,3]{Nunzia Lauriello\textsuperscript{\dag}}
\author[4]{Ilaria Cusano}
\author[4]{Marcello Ferranti}
\author[4]{Rossana Pasquino\textsuperscript{*}}
\author[1]{Safa Jamali\textsuperscript{*}}
\affil[1]{Department of Mechanical and Industrial Engineering,
Northeastern University, Boston, MA, USA}
\affil[2]{DISAT -- Department of Applied Science and Technology,
Politecnico di Torino, Corso Duca degli Abruzzi 24, 10129 Torino, Italy}
\affil[3]{Department of Physics ``A. Pontremoli'',
University of Milan, Via Celoria 16, 20133 Milan, Italy}
\affil[4]{DICMaPI -- Department of Chemical, Materials
and Industrial Production Engineering,
Universit\`a degli Studi di Napoli Federico II,
Piazzale Tecchio 80, 80125 Napoli, Italy}
\date{}
\begin{document}
\maketitle
\noindent\textsuperscript{\dag}These authors contributed equally to this work.\\
\noindent\textsuperscript{*}\textbf{Corresponding authors:}
rossana.pasquino@unina.it; s.jamali@northeastern.edu.\\
\noindent\textbf{Other author e-mail addresses:}
zolfaghari.s@northeastern.edu; nunzia.lauriello@unimi.it;
ilaria.cusano@unina.it; marcello.ferranti@unina.it.



\begin{abstract}
Wormlike micellar solutions exhibit formulation-dependent nonlinear startup responses, including stress overshoot, undershoot, and strain hardening, making prediction across compositions and shear rates challenging. We developed a multi-fidelity machine-learning framework to predict stress-growth curves in cetylpyridinium chloride (CPyCl)-based wormlike micelles from sparse experimental measurements. The framework combines a low-fidelity analytical model describing the overall startup response with high-fidelity experimental data capturing formulation-specific nonlinear features. This approach addresses the limited and uneven coverage of salt chemistry, concentration, and shear rate without relying solely on data-intensive neural-network training. The trained model predicts complete stress-growth responses at untested shear rates and enables extraction of strain-hardening characteristics from interpolated curves. By augmenting sparse rheological measurements, the framework provides a basis for exploring relationships among formulation variables, linear viscoelastic properties, and nonlinear startup behavior, while supporting more efficient characterization of wormlike micellar solutions.
\end{abstract}

\noindent\textbf{Keywords:} Wormlike micellar solutions; nonlinear
viscoelasticity; scientific machine learning; multi-fidelity modeling.

\section{Introduction}
Surfactants are amphiphilic molecules whose self-assembly behavior gives rise to a wide range of structural and rheological properties relevant to both fundamental research and industrial applications ~\cite{SHABAN}. In solution, these molecules spontaneously organize into ordered colloidal structures to minimize the exposure of their hydrophobic tails to water. The resulting self‑assembly morphology is primarily governed by the packing parameter, a geometrical quantity that characterizes the shape of the surfactant molecule, and dictates the curvature of the preferred aggregate ~\cite{Israel}.

Variations in external conditions such as concentration, temperature, and the presence of salts or other additives may induce structural rearrangements ~\cite{Dreiss2007, Ito2014}. In particular, aromatic counterions can promote micellar growth and the formation of wormlike aggregates ~\cite{Ito2014}. Such wormlike micelles (WLMs) are particularly interesting because they exhibit viscoelastic properties closely analogous to those of entangled polymeric systems. Although held together by weak, non‑covalent interactions, they can reptate like covalently bonded polymers while continuously breaking and reforming over time. This dynamic nature - simultaneously polymer‑like and transient — has led to WLMs being commonly referred to as “living polymers” ~\cite{Cates1987}.

Under startup shear flow, wormlike micellar systems can exhibit multiple nonlinear rheological phenomena, including stress overshoot, stress undershoot, shear banding, and strain hardening ~\cite{Rogers2014, Inoue2005, Pasquino2023, Gaudino2020, Divoux2016, Berret2006, Lopez}. These phenomena may occur on distinct characteristic timescales and are strongly influenced by micellar relaxation dynamics, finite extensibility, and scission processes ~\cite{CUSANO}. More specifically, the microstructural origin of transient strain hardening is still under debate. Historically, this behavior has been attributed either to micellar branching or to crosswise (electrostatic) contacts between worms, but neither explanation is fully convincing ~\cite{Inoue2005, Thurn}. However, branching alone is not a sufficient condition for strain hardening, since branched micellar systems that do not exhibit this response have been also reported ~\cite{Pasquino2023}. The contribution of crosswise electrostatic contacts to the observed stress evolution remains to be established quantitatively. It is, however, now well established that the accumulated deformation during start‑up shear drives the micelles toward their finite extensibility, which precedes their catastrophic breakage. Moreover, shear banding is generally associated with flow‑induced structural inhomogeneities and typically emerges after the stress overshoot, regardless of whether transient strain hardening is observed ~\cite{Lopez, Rassolov}.

Experimental studies have shown that whether this strain hardening behavior is observed, its onset and its magnitude vary significantly with surfactant chemistry, salt concentration, and applied deformation rate, suggesting a complex coupling between formulations and nonlinear flow response. Despite extensive experimental investigations, predictive understanding of startup stress-growth behavior across formulations remains limited. A major challenge is the sparse and uneven experimental coverage of formulation and shear-rate space. This limited coverage makes it even more challenging to map the occurrence of strain hardening and determine how its onset and intensity vary with composition and imposed shear rate. Furthermore, startup shear experiments are experimentally demanding and often distributed unevenly across formulation variables and shear conditions.

Recent advances in machine learning have provided new tools for constructing predictive models of complex physical systems directly from data. Neural networks are particularly effective at representing nonlinear relationships among material parameters, operating conditions, and system responses. Applications of scientific machine learning to complex fluids have demonstrated the potential of these methods to learn rheological behavior that is difficult to represent using conventional empirical models ~\cite{Lennon2023}. Nevertheless, conventional data-driven neural networks generally require large, high-quality, and sufficiently uniform training datasets. Their predictions may deteriorate in sparsely sampled or extrapolative regimes, and their data-driven nature can limit physical interpretability and robustness ~\cite{Carvalho2019, Karniadakis2021PIML}. Certainly in the case of WLMs nonlinear rheological response, purely data-driven machine learning approaches would require significantly larger and more uniformly sampled datasets to generalize reliably.

Physics-guided machine-learning methods seek to overcome the limitations of purely data-driven approaches by incorporating prior physical information into the learning architecture or optimization objective. Physics-informed neural networks embed governing equations into the loss function and have been extended to non-Newtonian fluid modeling, data-driven identification of viscoelastic constitutive models, fractional constitutive descriptions, and systems governed by nonlocal dynamics ~~\cite{Mahmoudabadbozchelou2022nnPINNs, Reyes2021PINN, Thakur2024ViscoelasticNet, Dabiri2023RhINN, Dabiri2025fPINNReview, Zolfaghari2026NonlocalPINN}. Other emerging approaches include rheology-informed graph neural networks and transformer-based architectures developed to learn complex-fluid constitutive behavior and flow responses ~~\cite{Mahmoudabadbozchelou2022DigitalRheometerTwins, saberi2025rheoformer}. A broader overview of data-driven and physics-informed approaches, their applications, and remaining challenges in rheology is provided by Mangal et al.~~\cite{Mangal2025Review}. These developments demonstrate the growing potential of machine learning in rheology. However, directly enforcing governing equations remains difficult when the mechanisms responsible for transient nonlinear behavior are incomplete, uncertain, or unavailable in closed form.

Multi-fidelity learning provides an alternative strategy for data-scarce problems in which the relevant physics is only partially known. In this framework, inexpensive low-fidelity information supplies dominant trends and broad parameter-space coverage, while limited high-fidelity measurements correct systematic discrepancies and introduce response features that simplified representations cannot capture ~\cite{Forrester2007MF, Perdikaris2017MF, Meng2020MPINN}. Multi-fidelity neural networks and related learning frameworks have demonstrated the ability to integrate information from different fidelity levels in rheology and other nonlinear physical systems ~\cite{Mahmoudabadbozchelou2021MFNN, Saadat2024MFNN, Boodaghidizaji2022MF, Howard2023MFDeepONet, Zolfaghari2026HierarchicalMF}. Despite this progress, the prediction of complete startup stress-growth responses across both strain-hardening and non-hardening wormlike micellar formulations remains largely unexplored. In particular, a successful model must represent stress growth, overshoot, undershoot, strain-hardening peaks, and post-peak relaxation across multiple formulations and imposed shear rates, without requiring a predefined classification of the hardening regime.

To address these limitations, we propose a multi-fidelity machine learning framework for predicting startup stress-growth behavior in cetylpyridinium chloride (CPyCl)-based wormlike micellar solutions using limited rheological data. The framework combines a low-fidelity analytical stress-growth model, capable of reproducing the general evolution of startup stress curves, with high-fidelity experimental measurements that capture formulation-specific nonlinear responses. The low-fidelity component provides physically informed structure and broad coverage across shear conditions, while the experimental data supply the nonlinear corrections associated with strain hardening and other formulation-dependent effects. By integrating these two sources of information, the proposed approach reduces the amount of experimental data required to generate predictive startup stress responses.

The trained model is used to predict stress-growth curves at untested shear rates and to interpolate nonlinear rheological features that cannot be directly extracted from sparse experiments alone. In particular, the framework enables estimation of strain-hardening metrics from reconstructed stress-growth responses, providing a route for augmenting incomplete rheological datasets. More broadly, this work represents a first step toward establishing predictive relationships between formulation chemistry, linear viscoelastic response, and nonlinear startup behavior in wormlike micellar systems through physics-guided machine learning. In the long term, such approaches may contribute to the development of predictive tools for the design and optimization of complex fluids with targeted nonlinear flow properties.

The remainder of this paper is organized as follows: Section 2 introduces the experimental dataset and the formulation of the startup stress-growth prediction problem. The multi-fidelity learning framework, including the analytical low-fidelity representation and the high-fidelity correction network, is, then, described in detail. Section 3 presents the predicted startup stress-growth responses and discusses the capability of the framework to reconstruct nonlinear rheological features across different formulations and imposed shear rates. Finally, the main conclusions and perspectives for future development are summarized in Section 4.

\section{Problem setup and Methodology}

Startup shear experiments provide direct access to the nonlinear stress-growth response of WLMs. However, constructing a complete experimental map of this response is challenging because each formulation must be tested over several imposed shear rates, and the resulting behavior can vary strongly with salt and surfactant type and their relative concentration. In particular, features such as stress overshoot, delayed stress growth, and strain hardening do not appear uniformly across all tested samples. As a result, the available measurements give detailed information for selected experimental conditions, but they do not provide a continuous description of the stress-growth response over the full formulation and imposed shear rates.

To address this limitation, we develop a multi-fidelity learning approach in which the experimental stress-growth curves are supported by an analytical approximation of the same response. The experimental measurements are treated as high-fidelity data because they provide the experimentally observed nonlinear rheological response of the tested systems. For each case, the high-fidelity target is the measured shear stress, $\sigma$, as a function of accumulated strain, $\gamma$, at a prescribed shear rate, $\dot{\gamma}$. These data include the formulation-specific response of CPyCl-based wormlike micellar solutions and therefore provide the reference behavior that the model is trained to reproduce.

The low-fidelity level is constructed from an analytical base--bump representation of the startup stress response. This representation is not intended to replace the experimental measurements. Instead, it provides a smooth approximate curve that captures the main structure of the stress-growth behavior. The base term describes the gradual increase of stress with strain, while the bump term allows localized deviations associated with overshoot or strain-hardening-like features. The parameters of this analytical response are obtained from polynomial regression models trained on fitted stress-growth parameters, allowing low-fidelity curves to be generated at both measured and intermediate shear-rate conditions.

The role of the multi-fidelity model is therefore to learn how the analytical low-fidelity estimate must be corrected in order to match the experimental high-fidelity response. In the present implementation, the low-fidelity curve is kept fixed and is supplied directly to the high-fidelity correction networks together with the strain, imposed shear rate, and formulation variables. The network then predicts the corrected high-fidelity stress response in logarithmic stress space. This formulation allows the model to use the analytical curve as a structural guide while retaining enough flexibility to capture nonlinear deviations present in the measured data.

\subsection{Experimental Dataset and Problem Formulation}
\label{sec:dataset_problem}

The dataset comprises both previously published and unpublished experimental measurements. The CPyCl--Diclo formulations containing $16.7\,\mathrm{mM}$ CPyCl were previously investigated in ~\cite{Cusano2024}. Previous studies have also reported startup shear responses for $100\,\mathrm{mM}$ CPyCl with $45$, $48$, or $50\,\mathrm{mM}$ NaSal in $100\,\mathrm{mM}$ NaCl brine ~\cite{Gaudino2020}, and for $100\,\mathrm{mM}$ CPyCl with either $68\,\mathrm{mM}$ NaSal or $52\,\mathrm{mM}$ Diclo ~\cite{Pasquino2023}. Measurements for the remaining formulations, comprising $200\,\mathrm{mM}$ CPyCl with $60$, $70$, $75$, $80$, or $90\,\mathrm{mM}$ Diclo, and $100\,\mathrm{mM}$ CPyCl with either $44$ or $50\,\mathrm{mM}$ Diclo or $73.1$ or $80\,\mathrm{mM}$ NaSal, are reported here for the first time. Materials, suppliers, sample preparation, and equilibration procedures for the CPyCl--Diclo solutions prepared in this work are described in ~\cite{Cusano2024}. Startup shear measurements were performed at $25\,^{\circ}\mathrm{C}$ using an ARES rheometer (TA Instruments, USA) equipped with a 50 mm-diameter cone-and-plate geometry. Before each startup test, the sample was allowed to rest for a period equal to three times its characteristic relaxation time. 
For each formulation, tests were conducted at multiple imposed shear rates, $\dot{\gamma}$, and the resulting shear stress, $\sigma$, was recorded as a function of the accumulated strain, $\gamma$. For selected experimental conditions, the imposed shear rate is also expressed in terms of the Weissenberg number, $Wi=\dot{\gamma}\tau_R$, where $\tau_R$ is the characteristic relaxation time of the corresponding formulation. Both quantities are reported in the captions of selected figures to facilitate comparison of the nonlinear responses across different formulations.

Each experimental case is defined by its formulation variables and imposed shear-rate condition. For the NaSal system, the formulation variables are the CPyCl and NaSal concentrations, whereas for the Diclo system, they are the CPyCl and Diclo concentrations. The measured response for each case is represented by a startup stress-growth curve,

\begin{equation}
    \sigma
    =
    \sigma\left(
    \gamma;
    \dot{\gamma},
    C_{\mathrm{CPyCl}},
    C_{\mathrm{salt}}
    \right),
    \label{eq:startup_response}
\end{equation}

where $C_{\mathrm{salt}}$ denotes either the NaSal or Diclo concentration, depending on the material system considered.

The formulation space covered by the high-fidelity dataset is summarized in Table~\ref{tab:experimental_formulations}. The formulations are grouped according to salt chemistry and the experimentally observed startup response within the investigated shear-rate range. Stress overshoots may occur in both hardening and non-hardening formulations; therefore, the presence of a stress maximum alone does not distinguish between the two responses. As shown in the table, the available measurements cover both strain-hardening and non-hardening formulations, but they are distributed nonuniformly across salt chemistry and concentration space.

\begin{table}[t]
    \centering
    \caption{Summary of the CPyCl-based wormlike micellar formulations included in the high-fidelity experimental dataset. Multiple salt concentrations listed in the same row correspond to the indicated CPyCl concentration. The response category indicates whether strain-hardening behavior was observed within the investigated range of imposed shear rates.}
    \label{tab:experimental_formulations}
    \renewcommand{\arraystretch}{1.2}
    \setlength{\tabcolsep}{10pt}
    \begin{tabular}{llcc}
        \toprule
        Salt system &
        Observed response &
        $C_{\mathrm{CPyCl}}$ (mM) &
        $C_{\mathrm{salt}}$ (mM) \\
        \midrule

        \multirow{5}{*}{Diclo}
        & \multirow{2}{*}{No strain hardening}
        & 200  & 60, 90 \\
        &      & 100  & 44, 52 \\
        \cmidrule(lr){2-4}

        & \multirow{3}{*}{Strain hardening}
        & 16.7 & 10, 10.5, 11 \\
        &      & 200  & 70, 75, 80 \\
        &      & 100  & 50 \\
        \midrule

        \multirow{2}{*}{NaSal}
        & No strain hardening
        & 100 & 45, 48, 50 \\
        \cmidrule(lr){2-4}

        & Strain hardening
        & 100 & 68, 73.1, 80 \\
        \bottomrule
    \end{tabular}
\end{table}

\begin{figure}[t]
    \centering

    \includegraphics[width=0.48\columnwidth]
    {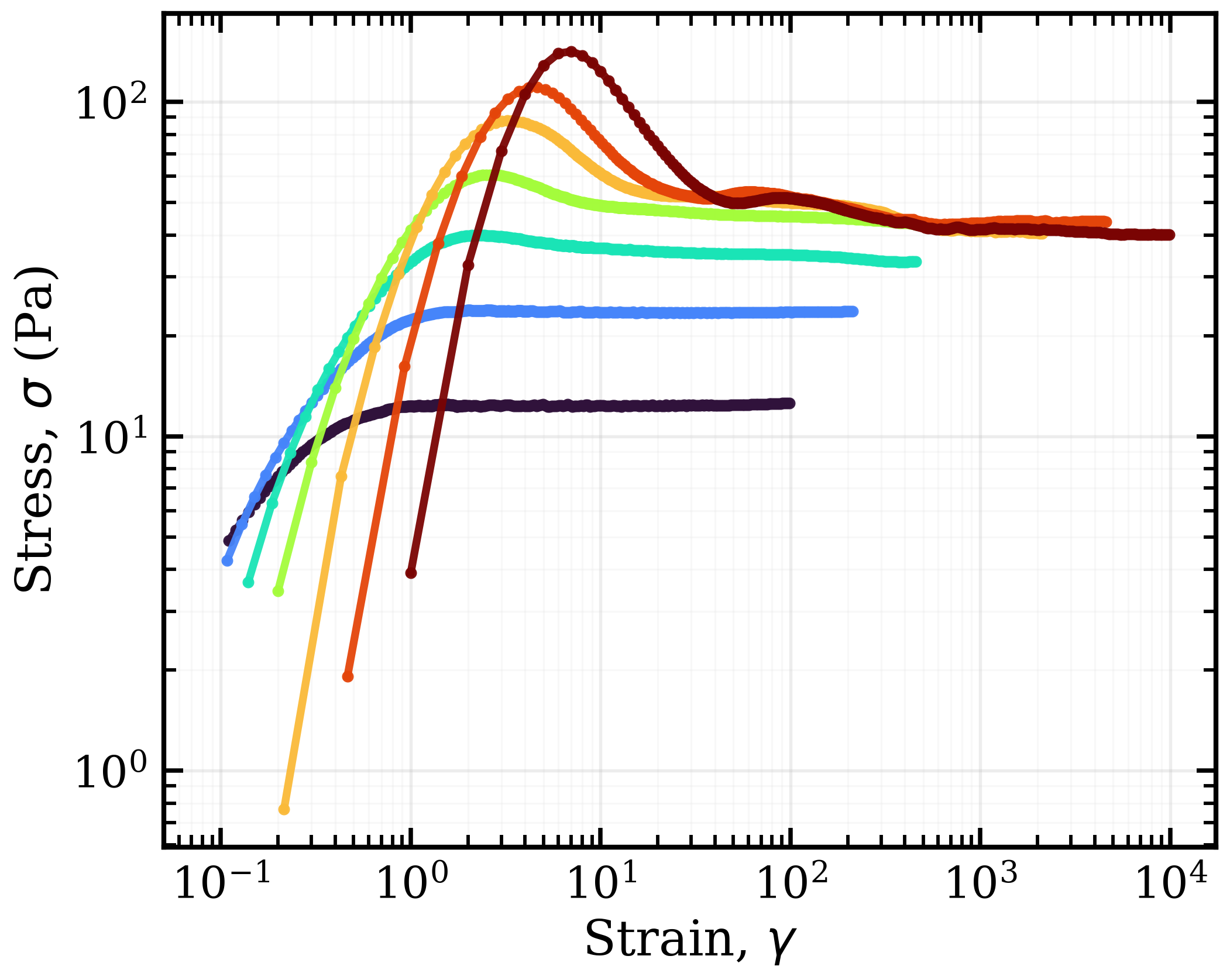}
    \includegraphics[width=0.48\columnwidth]
    {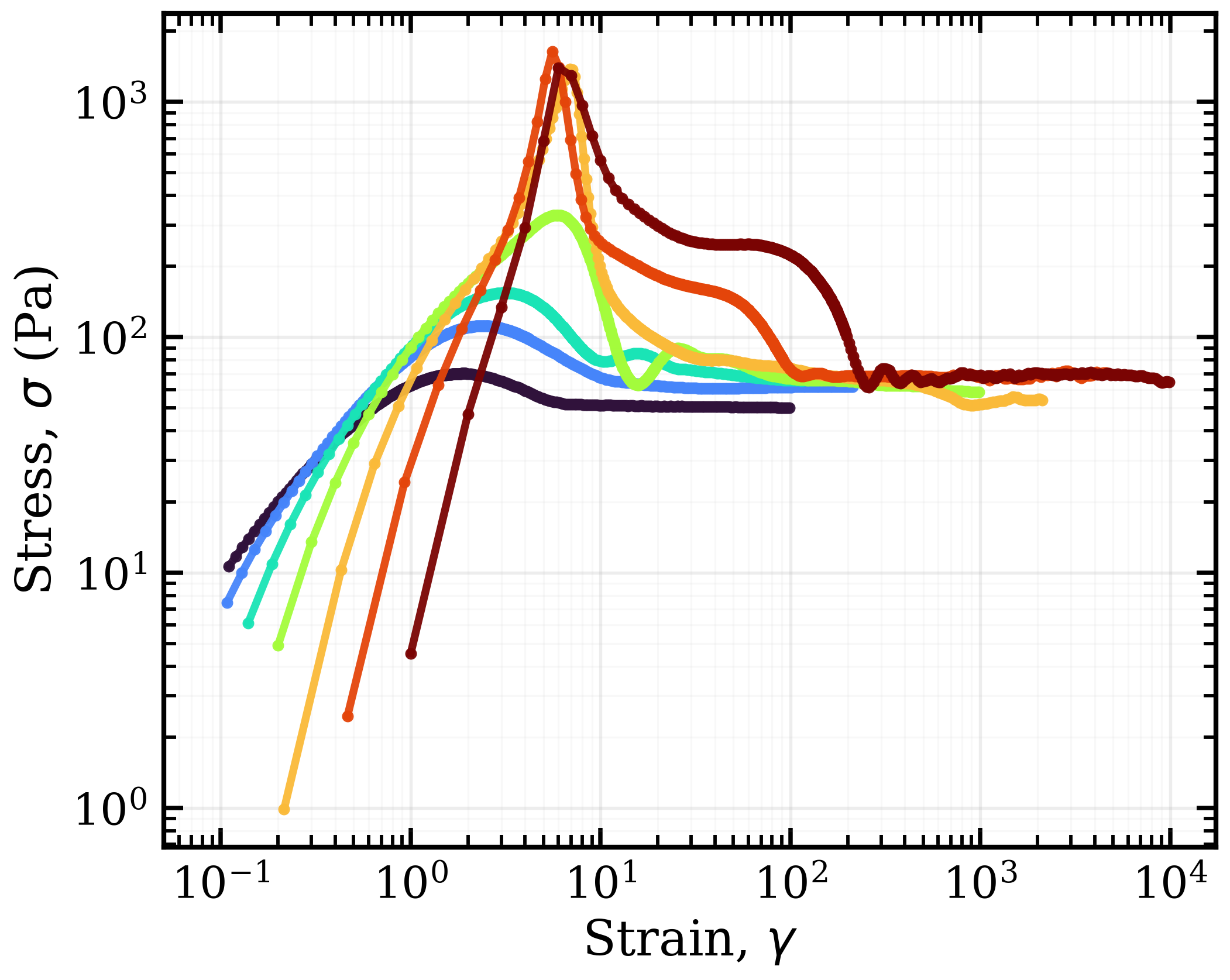}
    \caption{Experimental startup shear-stress responses for $200~\mathrm{mM}$ CPyCl with $60~\mathrm{mM}$ Diclo (left, no strain hardening) and $75~\mathrm{mM}$ Diclo (right, strain hardening). Both panels show $\dot{\gamma}=1$, $2.15$, $4.64$, $10$, $21.5$, $46.4$, and $100~\mathrm{s}^{-1}$ , with colors progressing from dark purple to dark red as $\dot{\gamma}$ increases [corresponding $Wi$: left $(0.147,\,0.316,\,0.681,\,1.468,\,3.156,\,6.811,\,14.678)$, $\tau_R=0.147~\mathrm{s}$; right $(0.681,\,1.465,\,3.161,\,6.813,\,14.648,\,31.612,\,68.129)$, $\tau_R=0.681~\mathrm{s}$].}

    \label{fig:diclo_hardening_comparison}
\end{figure}

The qualitative difference between the two experimentally observed response categories is illustrated in Figure~\ref{fig:diclo_hardening_comparison}. The figure compares two CPyCl--Diclo formulations having the same CPyCl concentration but different Diclo concentrations over their common shear-rate range. The formulation containing $60~\mathrm{mM}$ Diclo exhibits conventional stress growth followed by a stress overshoot or an approach toward a nearly steady response, without the development of pronounced strain hardening. In contrast, increasing the Diclo concentration to $75~\mathrm{mM}$ produces a strongly shear-rate-dependent hardening response characterized by rapid stress amplification, pronounced stress peaks, and subsequent post-peak relaxation. This comparison demonstrates that a change in salt concentration can produce substantial differences in the complete transient response, even when the CPyCl concentration and imposed shear rates are identical. This is well established in the literature ~\cite{Gaudino2015, Berret1995, Pasquino2023, Inoue2005}, as salt concentration can drive morphological transitions and drastically alter the linear viscoelastic response.

The high-fidelity dataset is constructed directly from these experimental startup shear curves. For each measured formulation and shear rate, the model inputs are the accumulated strain, imposed shear rate, CPyCl concentration, and salt concentration, while the target output is the experimentally measured shear stress. The resulting stress-growth responses exhibit formulation-dependent nonlinear behavior, including stress overshoot, delayed stress growth, pronounced strain hardening, and post-peak relaxation. Because these features do not occur uniformly across all formulations and shear-rate conditions, the experimental dataset provides detailed information at selected conditions but does not constitute a continuous map of the nonlinear response space.

The objective of the learning problem is to construct a surrogate model capable of predicting the complete startup stress-growth response for a specified formulation and imposed shear rate. In the first stage, the model is designed to reproduce the measured high-fidelity curves at experimentally tested shear rates and to predict responses at untested shear rates through interpolation and, where appropriate, extrapolation. This stage evaluates whether the model can learn the evolution of the complete transient response with $\dot{\gamma}$ within an individual formulation.

Beyond interpolation and extrapolation in shear-rate space, the broader objective is to enable prediction across formulation space by learning how changes in CPyCl concentration, salt concentration, imposed shear rate, and accumulated strain are connected to nonlinear startup behavior within each salt family. The resulting surrogate model provides a pathway for estimating responses between experimentally tested material cases and for augmenting sparsely sampled experimental datasets with model-generated stress-growth curves.

Once a complete stress-growth response has been predicted, nonlinear features associated with strain hardening can be extracted from the reconstructed curve. These features include the peak stress, peak strain, hardening-onset strain, and the range of imposed shear rates over which hardening occurs. The predicted nonlinear features can subsequently be combined with independently measured linear rheological properties to investigate possible relationships among material formulation, linear viscoelastic response, and nonlinear strain-hardening behavior.

\subsection{Multi-Fidelity Learning Framework}
\label{subsec:mf_model_formulation}

The present framework is formulated as a multi-fidelity surrogate model for predicting startup stress-growth curves of CPyCl-based wormlike micellar solutions. In this approach, the experimentally measured startup stress response is treated as the high-fidelity response, while a simplified analytical approximation of the same stress-growth behavior is used as the low-fidelity response. The low-fidelity model is not represented by a trainable neural network. Instead, it is constructed explicitly from an analytical base--bump function whose parameters are obtained from regression models trained on fitted experimental startup curves. This fixed low-fidelity response is then supplied as a structured input to the high-fidelity prediction networks.

\begin{figure}[t]
\centering
\includegraphics[width=0.94\textwidth]{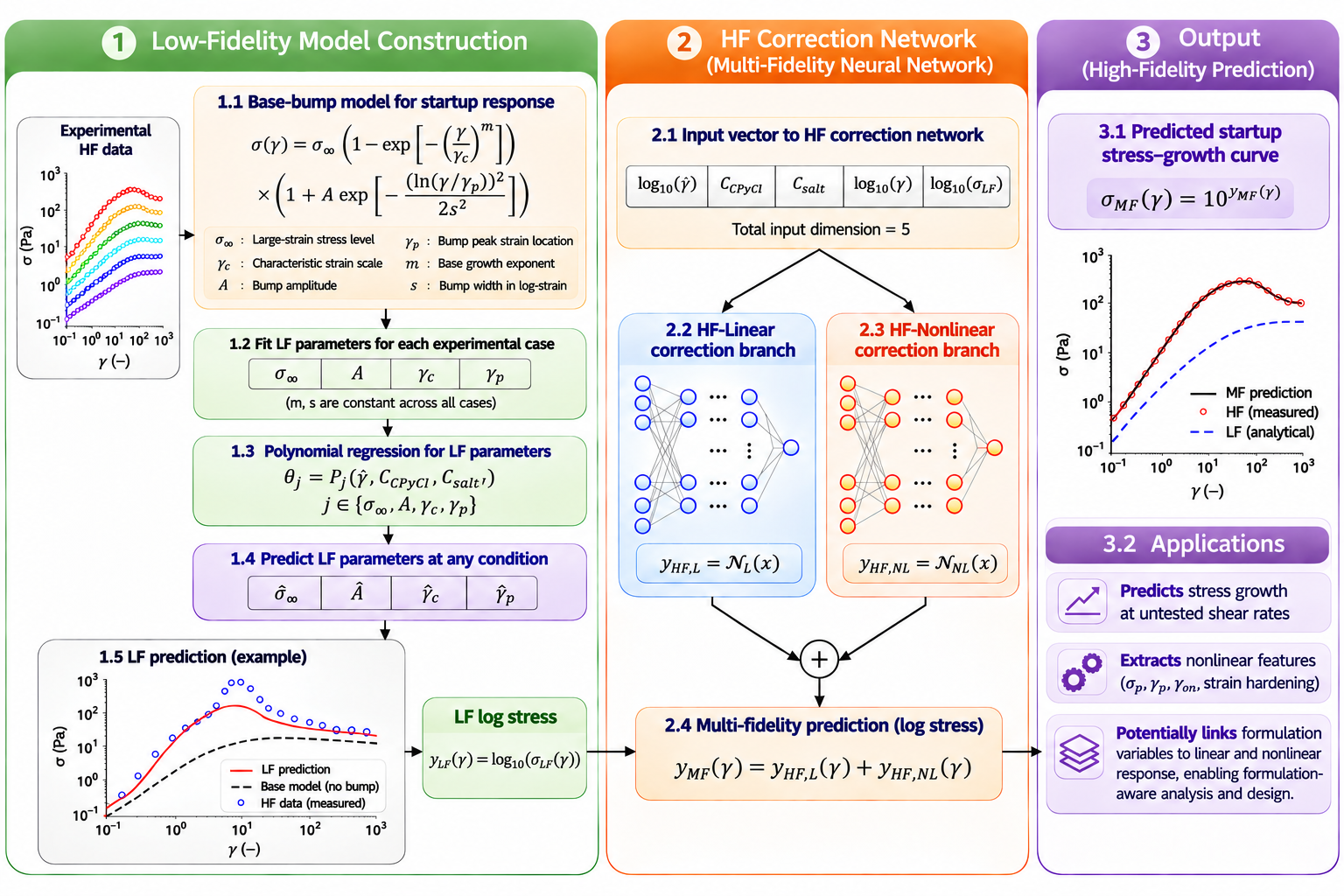}
\caption{
Schematic representation of the multi-fidelity learning framework used for predicting startup stress-growth curves in CPyCl-based wormlike micellar solutions. The workflow begins with construction of the low-fidelity model, where an analytical base–bump stress-growth representation is fitted to experimental stress–strain curves and the resulting parameters are modeled as functions of formulation variables and imposed shear rate. The predicted low-fidelity stress response is then supplied, together with the strain, shear rate, and formulation variables, to the high-fidelity prediction network. The high-fidelity network consists of linear and nonlinear branches whose outputs are summed to obtain the multi-fidelity prediction in logarithmic stress space. The final predicted stress-growth curve enables prediction over shear-rate and formulation conditions within each salt family, as well as extraction of nonlinear features such as peak stress, peak strain, onset strain, and strain-hardening-related metrics.
}
\label{fig:mufinn_architecture}
\end{figure}

The overall workflow of the proposed multi-fidelity learning framework is summarized in Fig.~\ref{fig:mufinn_architecture}. The framework consists of three connected stages. First, an analytical low-fidelity stress-growth representation is constructed from experimental startup curves using the base–bump model and polynomial regression models for the low-fidelity parameters. Second, the resulting low-fidelity prediction is used as a structured input to the high-fidelity correction network, which contains linear and nonlinear correction branches. Finally, the corrected logarithmic stress prediction is transformed back to stress space to obtain the predicted startup stress-growth curve and to enable extraction of nonlinear rheological features.

For each material system and imposed shear rate, the high-fidelity response is the experimentally measured shear stress,

\begin{equation}
    \sigma_{\mathrm{HF}}
    =
    \sigma_{\mathrm{exp}}
    \left(
    \gamma;
    \dot{\gamma},
    C_{\mathrm{CPyCl}},
    C_{\mathrm{salt}}
    \right),
\end{equation}

where $\gamma$ is the accumulated strain, $\dot{\gamma}$ is the imposed shear rate, $C_{\mathrm{CPyCl}}$ is the CPyCl concentration, and $C_{\mathrm{salt}}$ denotes the concentration of the added salt. Depending on the material family considered, $C_{\mathrm{salt}}$ corresponds to either the NaSal or Diclo concentration. Since the measured stress response spans different magnitudes over strain and shear-rate conditions, the learning problem is formulated in logarithmic stress space.

The low-fidelity stress response is defined using an analytical base--bump representation,

\begin{equation}
    \sigma_{\mathrm{LF}}(\gamma)
    =
    \sigma_{\infty}
    \left[
    1 - \exp\left(
    -\left(\frac{\gamma}{\gamma_c}\right)^m
    \right)
    \right]
    \left[
    1 + A
    \exp\left(
    -\frac{
    \left(\ln \gamma - \ln \gamma_p\right)^2
    }{2s^2}
    \right)
    \right].
    \label{eq:base_bump_lf}
\end{equation}

This functional form is chosen to provide a simple but flexible analytical description of startup stress-growth curves. Under startup shear, the stress typically evolves from a small initial value toward a larger strain-dependent response as deformation accumulates. This dominant buildup is represented by the base term,

\begin{equation}
    \sigma_{\infty}
    \left[
    1 - \exp\left(
    -\left(\frac{\gamma}{\gamma_c}\right)^m
    \right)
    \right],
\end{equation}

which describes a smooth monotonic increase of stress with strain. The parameter $\sigma_\infty$ sets the asymptotic stress level of the base response, while $\gamma_c$ defines the characteristic strain scale of stress buildup. Smaller values of $\gamma_c$ shift the rise toward lower strains, whereas larger values shift it toward higher strains. The exponent $m$ controls the shape of this transition.

However, many experimentally measured startup curves are not purely monotonic. Depending on formulation and imposed shear rate, the stress response may contain localized nonlinear features such as stress overshoot, delayed stress growth, or strain-hardening-like behavior over a finite range of strain. To represent these effects without introducing a more complex constitutive model, the monotonic base response is multiplied by a localized bump term,

\begin{equation}
    1 + A
    \exp\left[
    -\frac{
    \left(\ln \gamma - \ln \gamma_p\right)^2
    }{2s^2}
    \right].
\end{equation}

This term acts as a strain-localized amplification of the base stress response. The parameter $A$ controls the bump amplitude, with larger values corresponding to stronger localized enhancement of stress. The parameter $\gamma_p$ determines the strain at which this enhancement is centered and therefore identifies the approximate location of the localized nonlinear response. The parameter $s$ controls the width of the bump in logarithmic strain space; smaller values produce a more localized feature, while larger values produce a broader enhancement over a wider strain range.

The base--bump structure is therefore used because it separates the dominant stress-growth trend from localized nonlinear deviations. The base term captures the smooth buildup of stress that is common to startup shear responses, while the bump term provides enough flexibility to approximate overshoot or strain-hardening-like features when they are present. This makes the expression suitable as a low-fidelity representation: it is simple, smooth, and parameterized by physically interpretable quantities, but it is not expected to reproduce all experimental details. Instead, it provides a structured analytical approximation that guides the subsequent high-fidelity correction network.

Because the bump amplitude is constrained to be nonnegative, the analytical representation primarily describes localized stress amplification relative to the monotonic base response. Detailed undershoots, secondary oscillations, and formulation-specific post-peak relaxation are not expected to be reproduced exactly by the low-fidelity model. These discrepancies motivate the use of the high-fidelity neural-network component.

In the present implementation, the two shape parameters $m$ and $s$ are fixed for all cases,

\begin{equation}
    m = 1.25,
    \qquad
    s = 0.75.
\end{equation}

These parameters are kept constant to reduce the dimensionality of the fitting problem and to avoid over-parameterization of individual startup curves. Because the available experimental data are limited to a finite number of formulations and shear-rate conditions, allowing all six parameters in Eq.~\eqref{eq:base_bump_lf} to vary independently could lead to non-unique or unstable fits. Fixing $m$ and $s$ preserves a consistent base-growth and bump-width structure across all cases, while the remaining four parameters are allowed to vary with formulation and shear rate.

The four case-dependent low-fidelity parameters are therefore defined as

\begin{equation}
    \boldsymbol{\theta}_{\mathrm{LF}}
    =
    \left[
    \sigma_{\infty},
    \gamma_c,
    A,
    \gamma_p
    \right].
\end{equation}

These parameters are first obtained by fitting Eq.~\eqref{eq:base_bump_lf} to the available startup stress-growth curves. After this initial curve-fitting step, polynomial regression models are trained to describe how each low-fidelity parameter depends on the formulation variables and imposed shear rate. For both salt systems considered in this study, namely NaSal and Diclo, this mapping is written in the general form

\begin{equation}
    \theta_j
    =
    \mathcal{P}_j
    \left(
    C_{\mathrm{CPyCl}},
    C_{\mathrm{salt}},
    \dot{\gamma}
    \right),
    \qquad
    \theta_j \in
    \left\{
    \sigma_{\infty},
    \gamma_c,
    A,
    \gamma_p
    \right\},
    \label{eq:poly_param_general}
\end{equation}

where $C_{\mathrm{salt}}$ denotes the concentration of the added salt component, corresponding to either $C_{\mathrm{NaSal}}$ or $C_{\mathrm{Diclo}}$ depending on the material system. The function $\mathcal{P}_j$ denotes a fourth-degree polynomial regression model for the $j$th low-fidelity parameter. Separate polynomial parameter models are constructed for the NaSal and Diclo formulation families. The polynomial inputs are the CPyCl concentration, the concentration of the corresponding salt, and the imposed shear rate. The resulting parameter predictions are constrained during post-processing to satisfy $\hat{\sigma}_{\infty}>0$, $\hat{\gamma}_c>0$, $\hat{A}\geq0$, and $\hat{\gamma}_p>0$ before evaluation of the analytical stress response. In this way, the fitted low-fidelity parameters are converted into smooth functions of composition and shear rate, allowing analytical low-fidelity stress-growth curves to be generated at measured and intermediate conditions within the explored parameter space.

Once the polynomial parameter models are constructed, they are used to generate the low-fidelity curve for any specified input condition. The predicted low-fidelity parameter vector is written as

\begin{equation}
    \hat{\boldsymbol{\theta}}_{\mathrm{LF}}
    =
    \left[
    \hat{\sigma}_{\infty},
    \hat{\gamma}_c,
    \hat{A},
    \hat{\gamma}_p
    \right],
\end{equation}

and these predicted parameters are substituted into Eq.~\eqref{eq:base_bump_lf} to obtain the fixed low-fidelity stress response,

\begin{equation}
    \sigma_{\mathrm{LF}}^{\mathrm{fixed}}
    =
    \sigma_{\mathrm{LF}}
    \left(
    \gamma;
    \hat{\sigma}_{\infty},
    \hat{\gamma}_c,
    \hat{A},
    \hat{\gamma}_p,
    m,
    s
    \right).
    \label{eq:fixed_lf}
\end{equation}

The high-fidelity correction model then uses this fixed low-fidelity prediction as an additional structured input. The input vector to the correction network is defined as

\begin{equation}
    \mathbf{x}
    =
    \left[
    \log_{10}(\gamma),
    \log_{10}(\dot{\gamma}),
    C_{\mathrm{CPyCl}},
    C_{\mathrm{salt}},
    \log_{10}
    \left(
    \sigma_{\mathrm{LF}}^{\mathrm{fixed}}
    \right)
    \right].
    \label{eq:mf_input_vector}
\end{equation}

Before network training, each component of the input vector is standardized using the mean and standard deviation calculated exclusively from the training data. For the $k$th input component, the standardized value is defined as

\begin{equation}
    \widetilde{x}_k
    =
    \frac{x_k-\mu_k}{s_k},
    \label{eq:input_standardization}
\end{equation}

where $\mu_k$ and $s_k$ denote the mean and standard deviation of the $k$th input component in the training dataset, respectively. The same training-derived normalization parameters are used to standardize the inputs when evaluating untested conditions.

The model output is the corrected high-fidelity stress response in logarithmic form,

\begin{equation}
    y_{\mathrm{MF}}
    =
    \log_{10}
    \left(
    \sigma_{\mathrm{MF}}
    \right).
\end{equation}

Unlike a standard multi-fidelity neural network in which the low-fidelity response is generated by a trainable LF network, the present implementation keeps the analytical low-fidelity response fixed. The neural network is therefore used only to learn the correction from the analytical low-fidelity estimate to the experimental high-fidelity response. The high-fidelity prediction is decomposed into a linear and a nonlinear correction branch,

\begin{equation}
    y_{\mathrm{MF}}
    =
    \mathcal{N}_{L}(\mathbf{x})
    +
    \mathcal{N}_{NL}(\mathbf{x}),
    \label{eq:fixed_lf_mf_prediction}
\end{equation}

where $\mathcal{N}_{L}$ denotes the lower-complexity correction branch and $\mathcal{N}_{NL}$ denotes the nonlinear correction branch. The linear branch represents the dominant global correction between the low- and high-fidelity responses, while the nonlinear branch provides additional flexibility to capture localized and higher-order discrepancies associated with nonlinear startup features. Both branches receive the same standardized input vector, including the fixed low-fidelity stress. Their scalar outputs are summed to predict the logarithmic high-fidelity response. Thus, the model does not impose an explicit additive residual correction to the low-fidelity curve; instead, it learns the high-fidelity stress as a function conditioned on the analytical low-fidelity estimate.

In practice, the nonlinear branch consists of six (6) hidden layers with 64 neurons per layer and hyperbolic-tangent activation functions. This branch provides the flexibility required to represent formulation-dependent nonlinear features, including sharp stress growth, localized strain-hardening peaks, and post-peak relaxation. The lower-complexity branch consists of two hidden layers with 32 neurons per layer and linear activation functions. Because compositions of linear layers remain linear mappings, this branch represents the dominant affine relationship between the standardized inputs and the logarithmic high-fidelity stress. The trainable parameters are determined by minimizing a regularized mean-squared-error objective in logarithmic stress space,

\begin{equation}
    \mathcal{L}
    =
    \frac{1}{N}
    \sum_{i=1}^{N}
    \left[
    \log_{10}
    \left(
    \sigma_{\mathrm{exp},i}
    \right)
    -
    y_{\mathrm{MF},i}
    \right]^2
    +
    \frac{\lambda_{NL}}{2}
    \left\|
    \mathbf{w}_{NL}
    \right\|_2^2
    +
    \frac{\lambda_L}{2}
    \left\|
    \mathbf{w}_L
    \right\|_2^2.
    \label{eq:fixed_lf_loss}
\end{equation}

where $N$ is the number of high-fidelity training points, $\mathbf{w}_{NL}$ and $\mathbf{w}_{L}$ denote the trainable parameters of the nonlinear and linear branches, respectively, and $\lambda_{NL}$ and $\lambda_L$ are the corresponding regularization coefficients. The first term enforces agreement between the predicted and experimental startup stress responses, while the regularization terms penalize large network parameters and reduce overfitting.

Based on holdout validation, both coefficients are set to $\lambda_{NL}=10^{-5}$ and $\lambda_L=10^{-6}$. The larger coefficient assigned to the nonlinear branch reflects its greater depth, parameter count, and representational capacity, which make it more susceptible to overfitting. The substantially smaller coefficient applied to the linear branch introduces only a weak penalty while allowing this lower-capacity component to represent the dominant global relationship between the low- and high-fidelity responses.

The model was trained using the Adam optimizer for a maximum of 200,000 iterations. A piecewise-constant learning-rate schedule was employed, with learning rates of $3\times10^{-3}$, $1\times10^{-3}$, and $5\times10^{-4}$ and transitions after 5,000 and 12,000 iterations, respectively. Early stopping was activated after a minimum of 20,000 iterations. The loss was evaluated every 500 iterations, and training was terminated when no relative improvement greater than $10^{-4}$ was observed over 40 consecutive evaluations, corresponding to 20,000 iterations without sufficient improvement. The regularization coefficients were selected based on holdout validation and were set to $\lambda_{NL}=10^{-5}$ and $\lambda_L=10^{-6}$. All normalization statistics are calculated exclusively from the corresponding training subset. The same training-derived means and standard deviations are subsequently used to standardize inputs under untested conditions.

This fixed low-fidelity formulation provides two main advantages. First, the analytical base--bump curve gives the correction network a structured estimate of the startup stress response, including the dominant stress buildup and possible localized nonlinear features. Second, because the low-fidelity response is generated through parameterized regression models, it can be extended to conditions within the explored experimental range. The correction network can therefore be used to predict stress-growth curves at untested shear-rate conditions and, with sufficient coverage of formulation space, to examine how nonlinear strain-hardening-related features vary with salt type, salt concentration, surfactant concentration, and the measured material response.

\begin{figure}[t]
    \centering
    \begin{subfigure}[t]{0.25\textwidth}
        \centering
        \includegraphics[width=\linewidth]
        {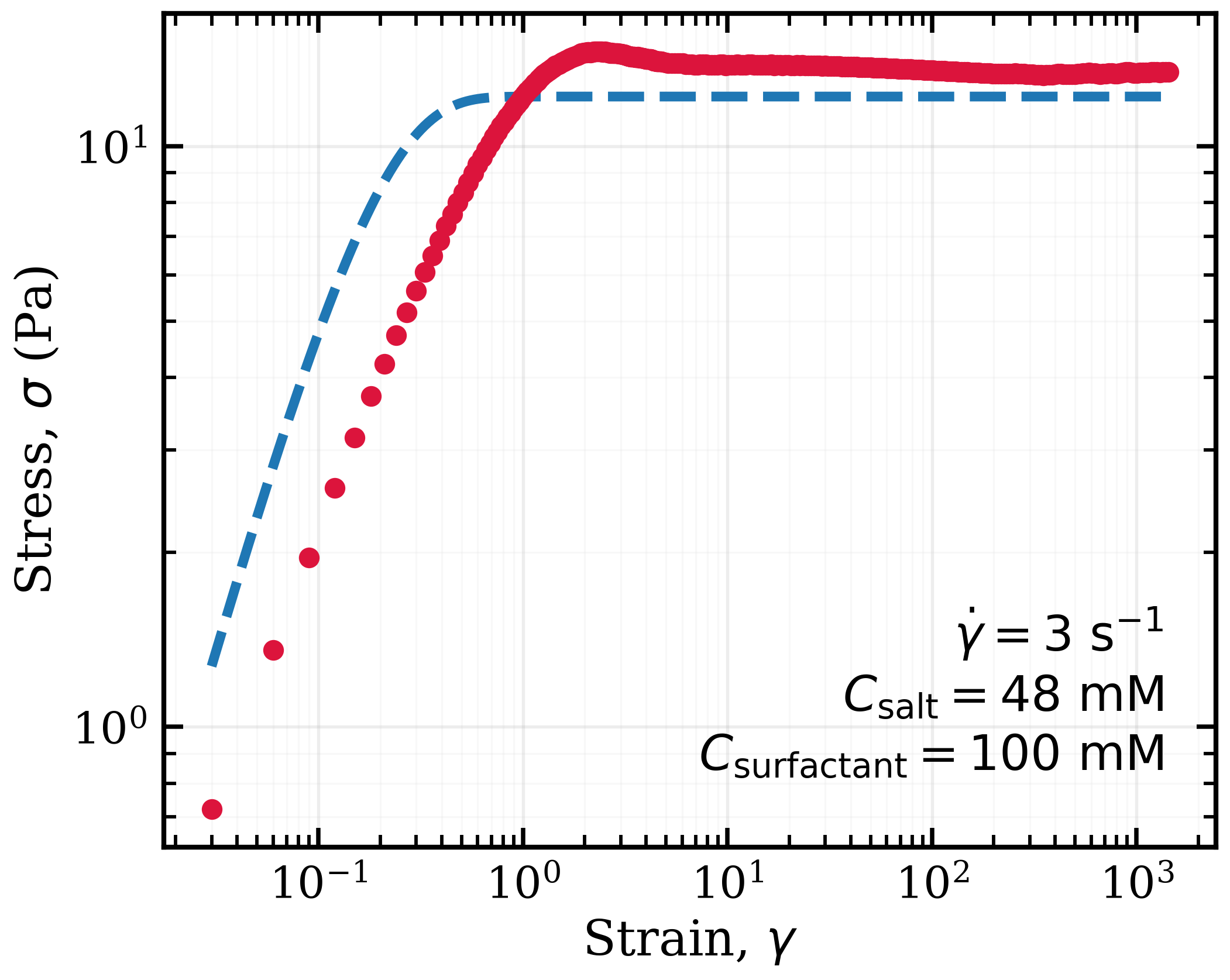}
    \end{subfigure}\hfill
    %
    \begin{subfigure}[t]{0.22\textwidth}
        \centering
        \includegraphics[
            width=\linewidth,
            trim={30 0 0 0},
            clip
        ]{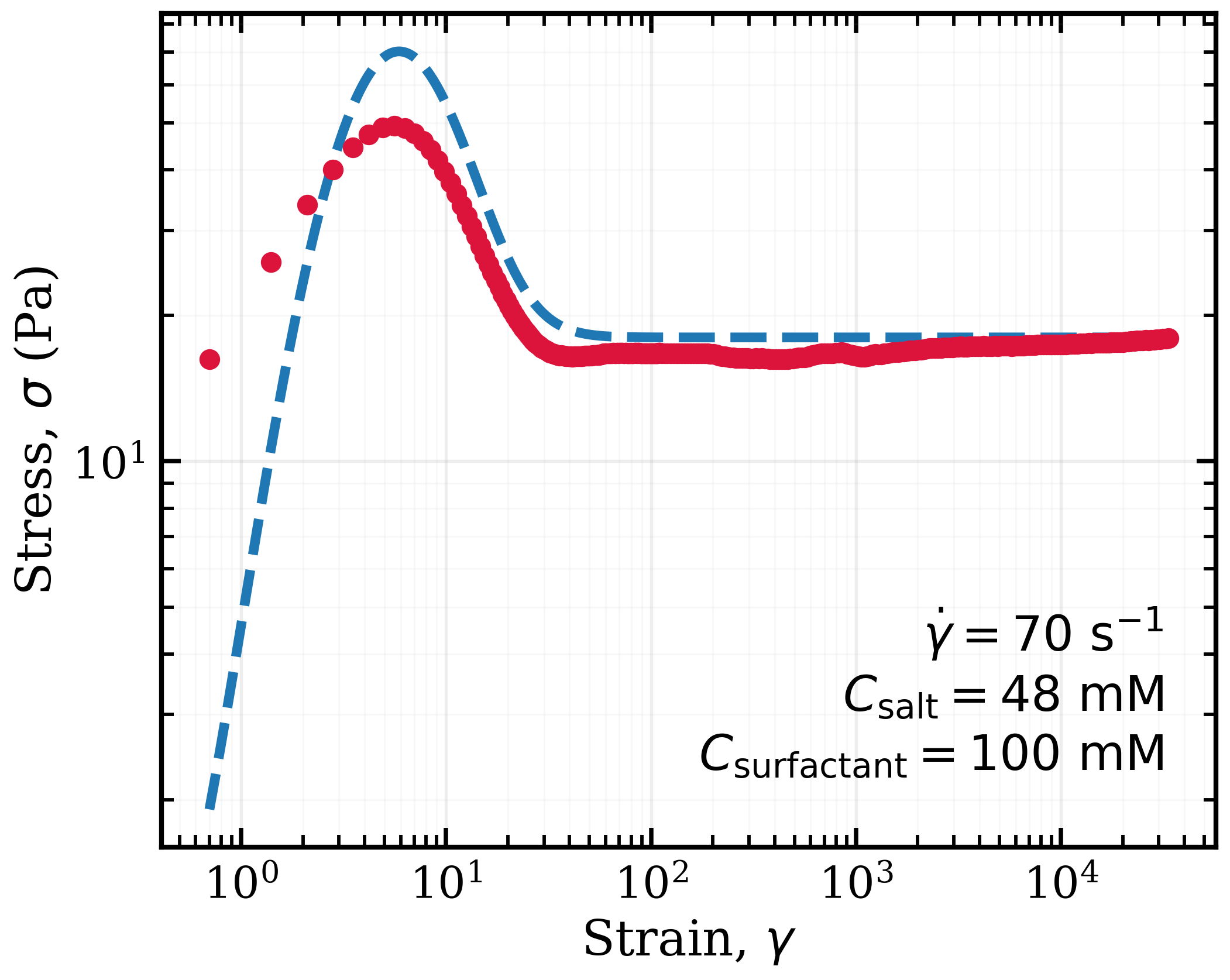}
    \end{subfigure}\hfill
    \begin{subfigure}[t]{0.22\textwidth}
        \centering
        \includegraphics[
            width=\linewidth,
            trim={30 0 0 0},
            clip
        ]{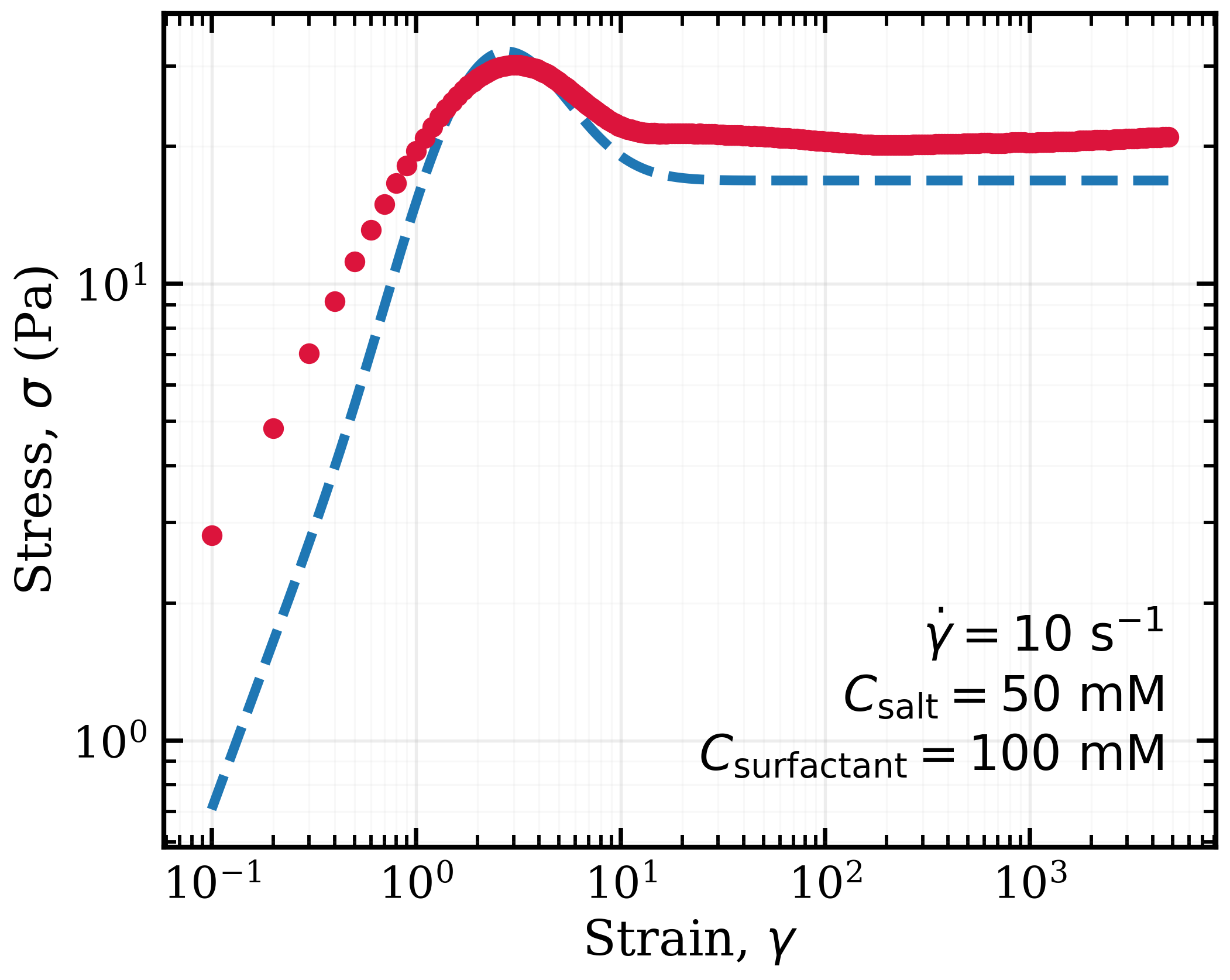}
    \end{subfigure}\hfill
    \begin{subfigure}[t]{0.22\textwidth}
        \centering
        \includegraphics[
            width=\linewidth,
            trim={30 0 0 0},
            clip
        ]{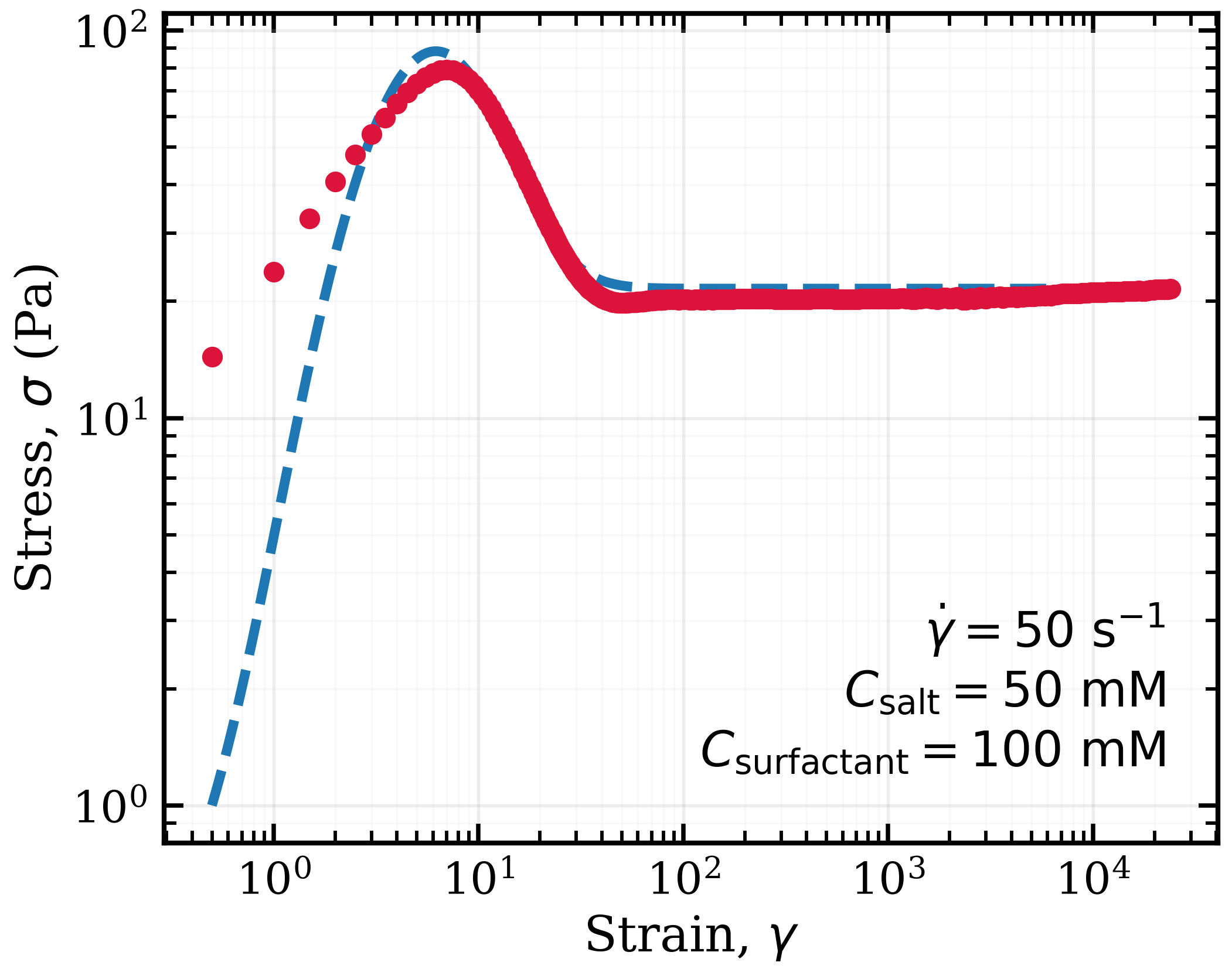}
    \end{subfigure}

    \vspace{0.8em}
    \begin{subfigure}[t]{0.25\textwidth}
        \centering
        \includegraphics[width=\linewidth]
        {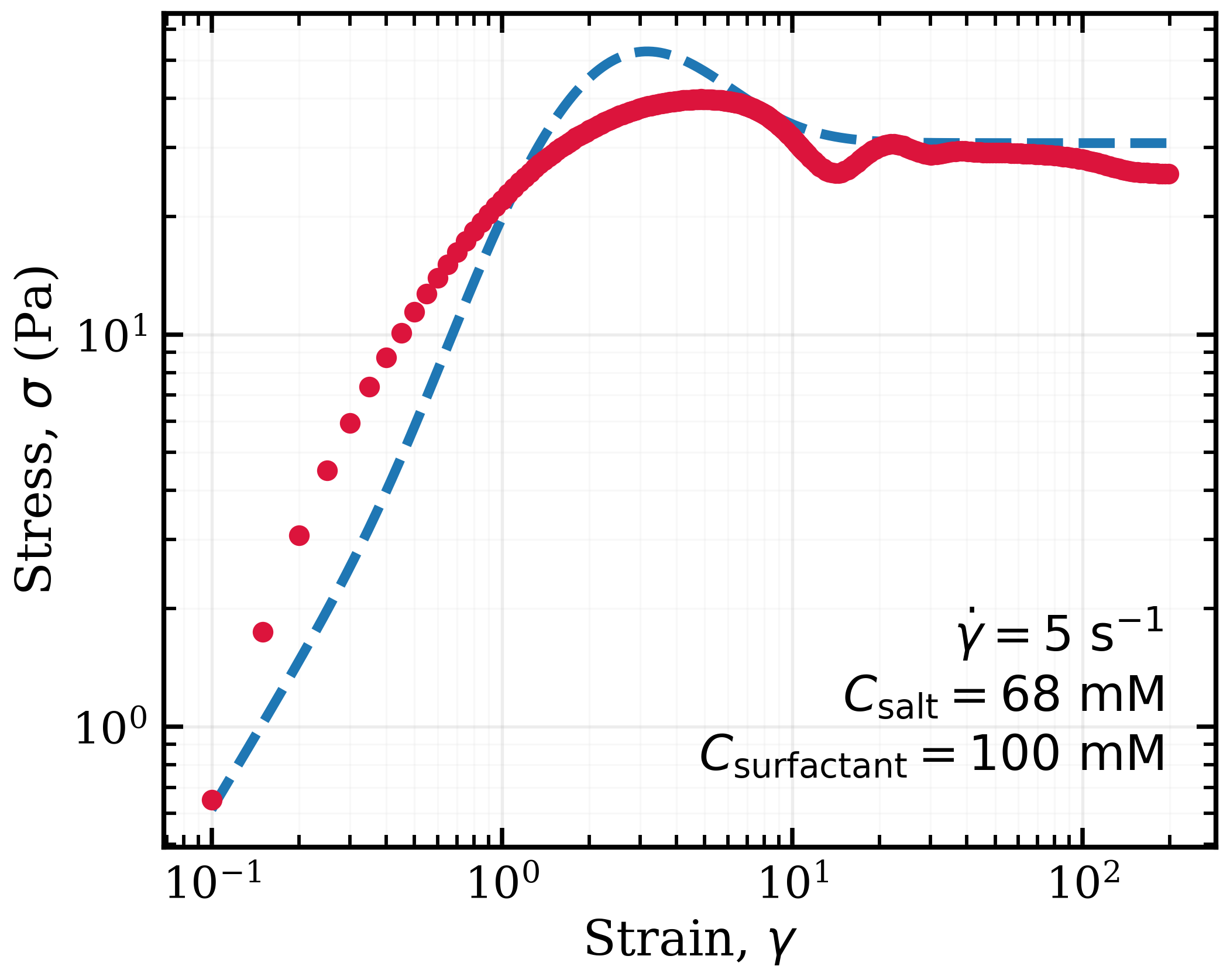}
    \end{subfigure}\hfill
    %
    \begin{subfigure}[t]{0.22\textwidth}
        \centering
        \includegraphics[
            width=\linewidth,
            trim={30 0 0 0},
            clip
        ]{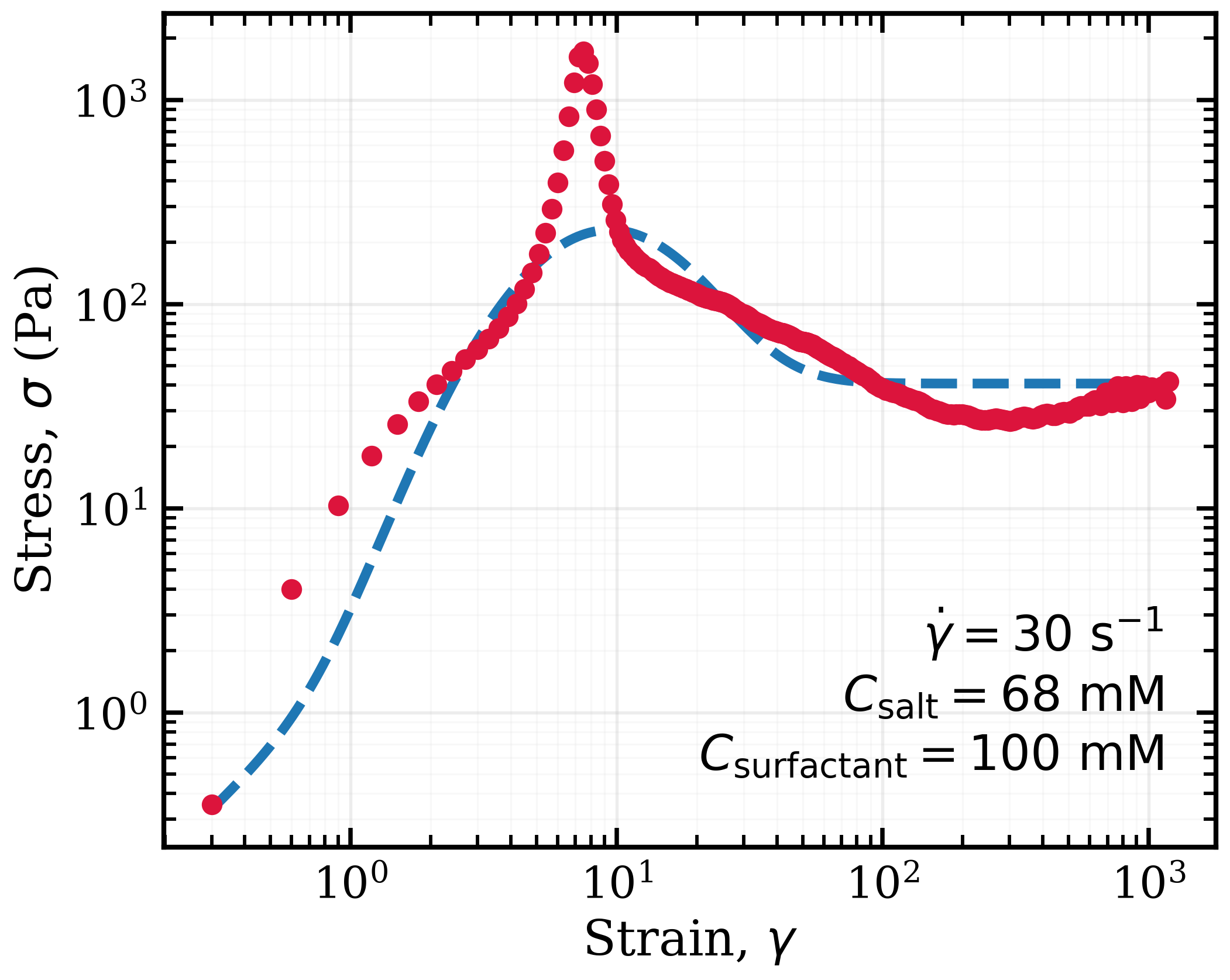}
    \end{subfigure}\hfill
    \begin{subfigure}[t]{0.22\textwidth}
        \centering
        \includegraphics[
            width=\linewidth,
            trim={30 0 0 0},
            clip
        ]{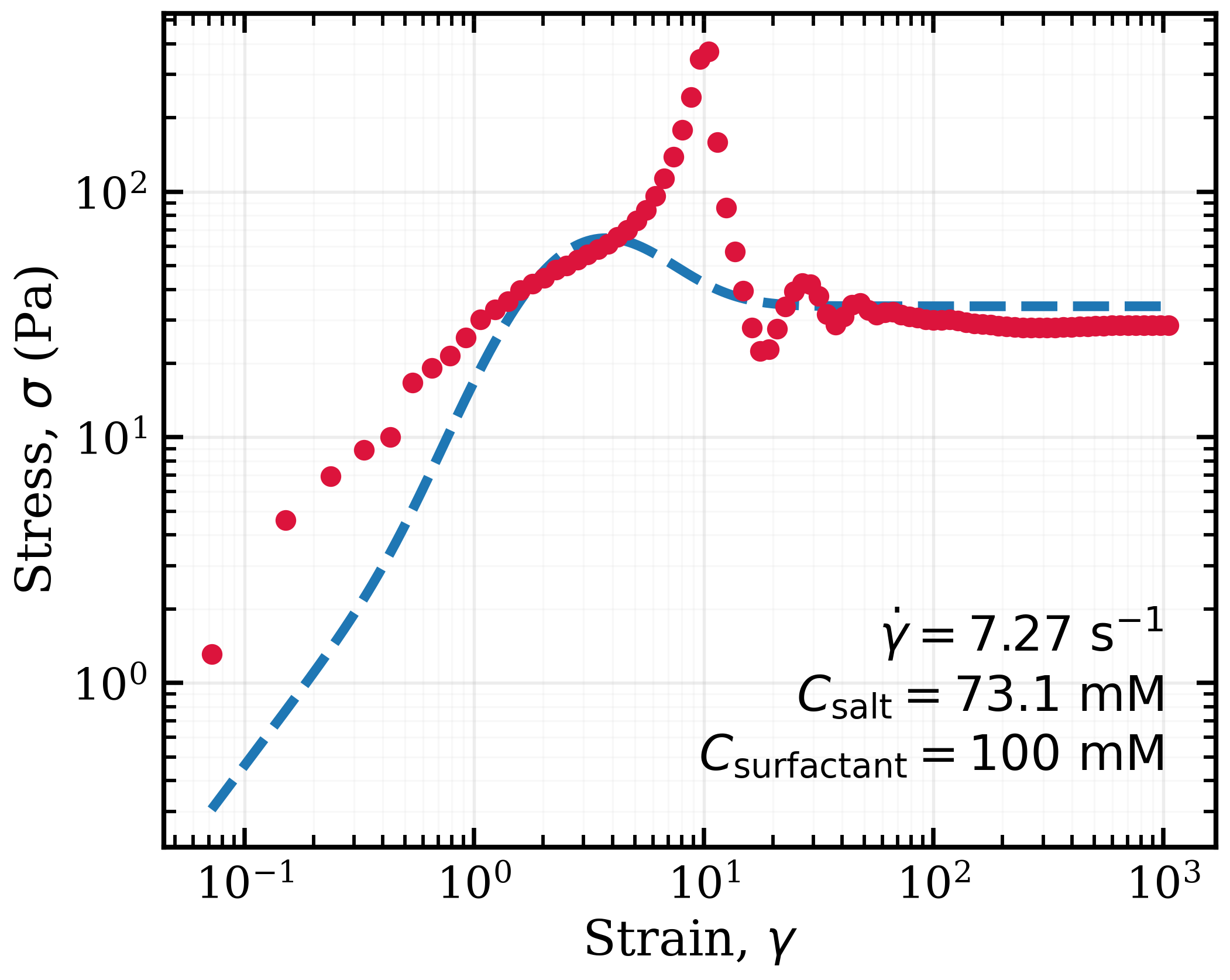}
    \end{subfigure}\hfill
    \begin{subfigure}[t]{0.22\textwidth}
        \centering
        \includegraphics[
            width=\linewidth,
            trim={30 0 0 0},
            clip
        ]{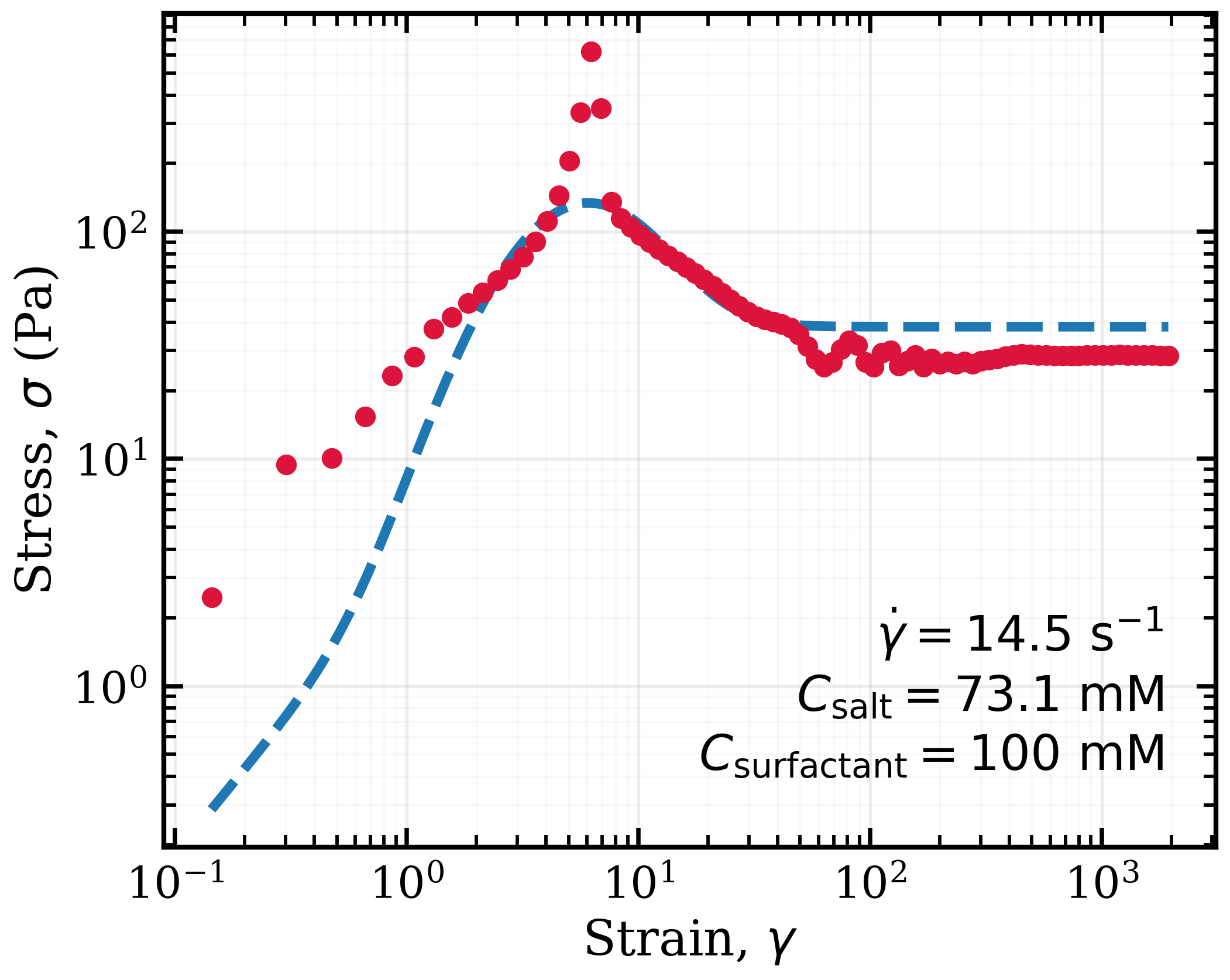}
    \end{subfigure}

    \caption{
    Comparison of high-fidelity experimental startup shear-stress responses (red symbols) and the corresponding low-fidelity analytical predictions (blue dashed curves) for representative CPyCl--NaSal formulations. The upper row shows systems without strain hardening, whereas the lower row shows strain-hardening systems. Within each formulation, responses at two imposed shear rates are presented.}
    \label{fig:hf_lf_nasal}
\end{figure}

\section{Results and Discussion}

To evaluate the predictive capability of the multi-fidelity framework, we first examined whether it could reconstruct complete startup stress-growth responses at shear rates for which no high-fidelity measurements were provided during training. This capability is important because experimentally characterizing every formulation over a densely sampled range of shear rates can be costly and time consuming. If reliable predictions can be obtained between a limited number of measured shear rates, the model can be used to interpolate the response at intermediate conditions and thereby reduce the number of experiments required to characterize the shear-rate dependence of a formulation. Extrapolation is also of practical interest, particularly toward high shear rates, where startup measurements may become difficult or unavailable because of the short duration of the transient response, instrument inertia, limited temporal resolution, or the onset of flow instabilities (e.g. elastic instabilities and edge fracture). The purpose of the present holdout analysis was therefore to establish, under controlled conditions, whether the multi-fidelity framework could recover missing responses both within and at the boundaries of the experimentally sampled shear-rate range.

The CPyCl--NaSal and CPyCl--Diclo systems were evaluated independently. Within each salt family, the model was trained using the remaining high-fidelity measurements from the available formulations together with the corresponding low-fidelity information. Complete high-fidelity curves associated with selected formulation--shear-rate conditions were deliberately withheld before training. Although these experimental curves were available in the original dataset, they were hidden from the model and retained exclusively as ground truth for validation. This procedure emulates the practical situation in which experiments are available at only a limited number of shear rates and the response must be estimated at the unmeasured conditions.

For every withheld condition, all high-fidelity stress values belonging to the corresponding startup curve were removed. Thus, the model did not observe any portion of the target experimental response during training. The fixed low-fidelity curve at the target condition remained available as a structured approximation of the general startup behavior, while the multi-fidelity network learned its correction using the remaining high-fidelity data. The resulting prediction was then evaluated against the complete withheld experimental curve. This is a more demanding test than randomly removing individual points from a curve, because the network must reconstruct the entire transient response at an excluded shear rate rather than fill small gaps within an otherwise observed curve.

The validation was performed in three progressively more challenging stages. In the first stage, one complete shear-rate curve was removed from an individual formulation at a time. Intermediate shear rates were initially withheld to evaluate interpolation. Under these conditions, high-fidelity curves at both lower and higher shear rates remained available for the target formulation. The model was therefore required to infer the response within a gap in the experimentally sampled shear-rate space. This test directly addresses whether a formulation can be characterized experimentally at a smaller number of selected shear rates, while the MFNN is subsequently used to estimate the responses at intermediate conditions.

Interpolation was evaluated for the complete stress-growth response rather than for a single rheological quantity. Successful prediction therefore required simultaneous reconstruction of the initial stress buildup, stress overshoot, possible strain-hardening peak, peak magnitude and location, post-peak relaxation, and large-strain response. This distinction is particularly important for strain-hardening formulations because the variation with $\dot{\gamma}$ cannot generally be represented only by a smooth change in stress magnitude. The complete shape of the transient curve, including the development or disappearance of localized nonlinear features, can also change substantially with shear rate.

The lowest and highest available shear rates were then withheld to evaluate extrapolation. Once a boundary curve was removed, the target shear rate lay outside the interval represented by the remaining high-fidelity measurements for that formulation. Unlike interpolation, the model could no longer use experimental curves on both sides of the target condition. It instead had to combine the shear-rate trend learned from the available high-fidelity data with the structured response supplied by the low-fidelity model.

The high-shear-rate holdout is particularly relevant to the intended application of the framework. Experiments at increasingly large $\dot{\gamma}$ may be challenging because the relevant transient occurs over a short time scale and may be affected by instrumental or flow-related limitations. By deliberately withholding the highest experimentally available shear rate, the analysis provides a controlled assessment of whether the model can reconstruct a response beyond the high-fidelity training interval. Agreement with the withheld measurement increases confidence that the framework may be used for limited extrapolation toward high-shear-rate conditions where direct experiments are unavailable or difficult to perform. Such use nevertheless remains subject to the distance between the prediction condition and the experimentally supported domain.

The single-shear-rate holdout procedure was repeated independently for the selected shear rates of each formulation. A separate MFNN was trained for every excluded condition. Consequently, each displayed prediction at a particular value of $\dot{\gamma}$ was generated by a model for which the complete corresponding high-fidelity curve had been removed. 

\begin{figure}[t]
    \centering
    \begin{minipage}[t]{0.49\textwidth}
        \centering
        \includegraphics[width=\linewidth]
        {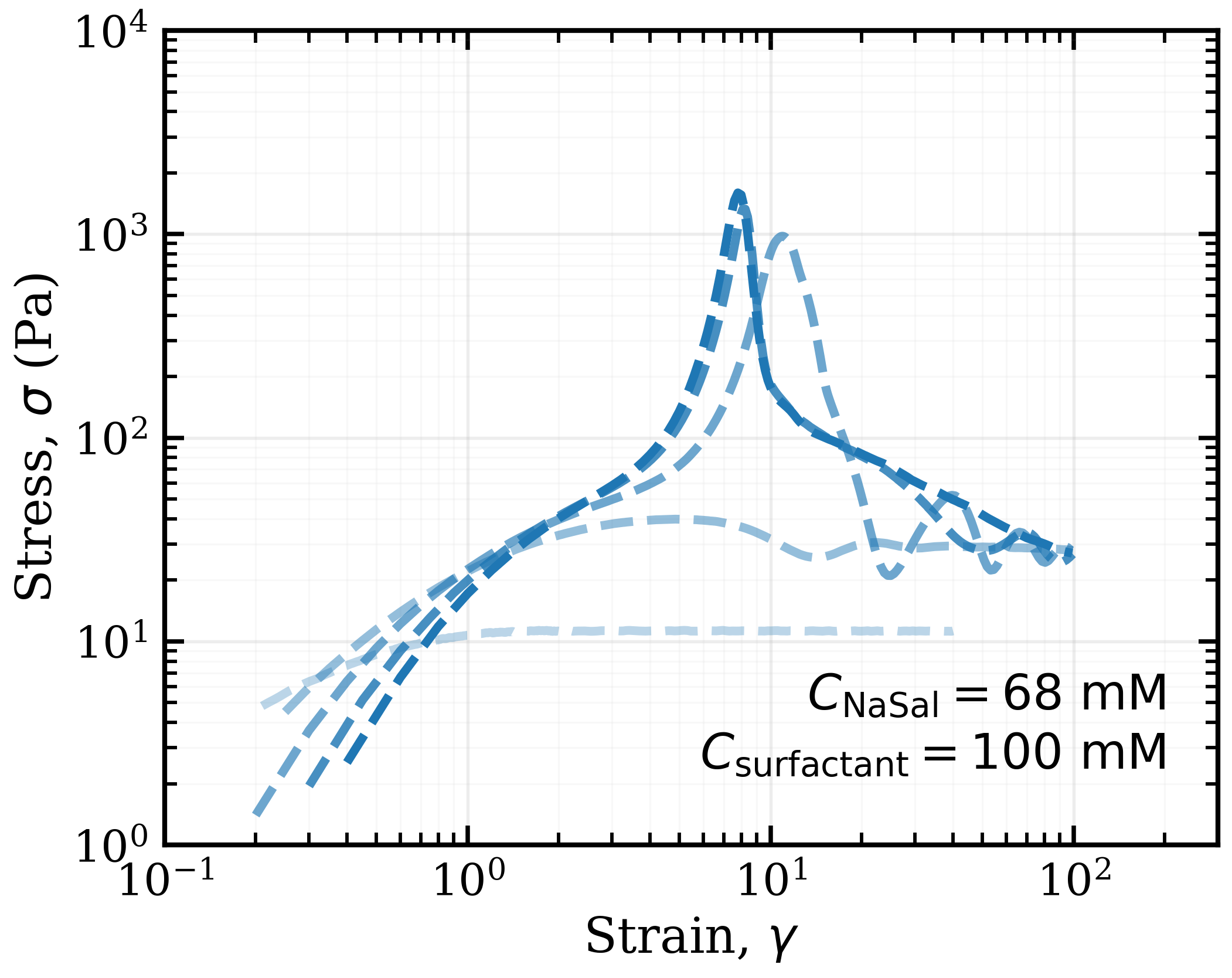}
    \end{minipage}%
    \hfill%
    \begin{minipage}[t]{0.49\textwidth}
        \centering
        \includegraphics[width=\linewidth]
        {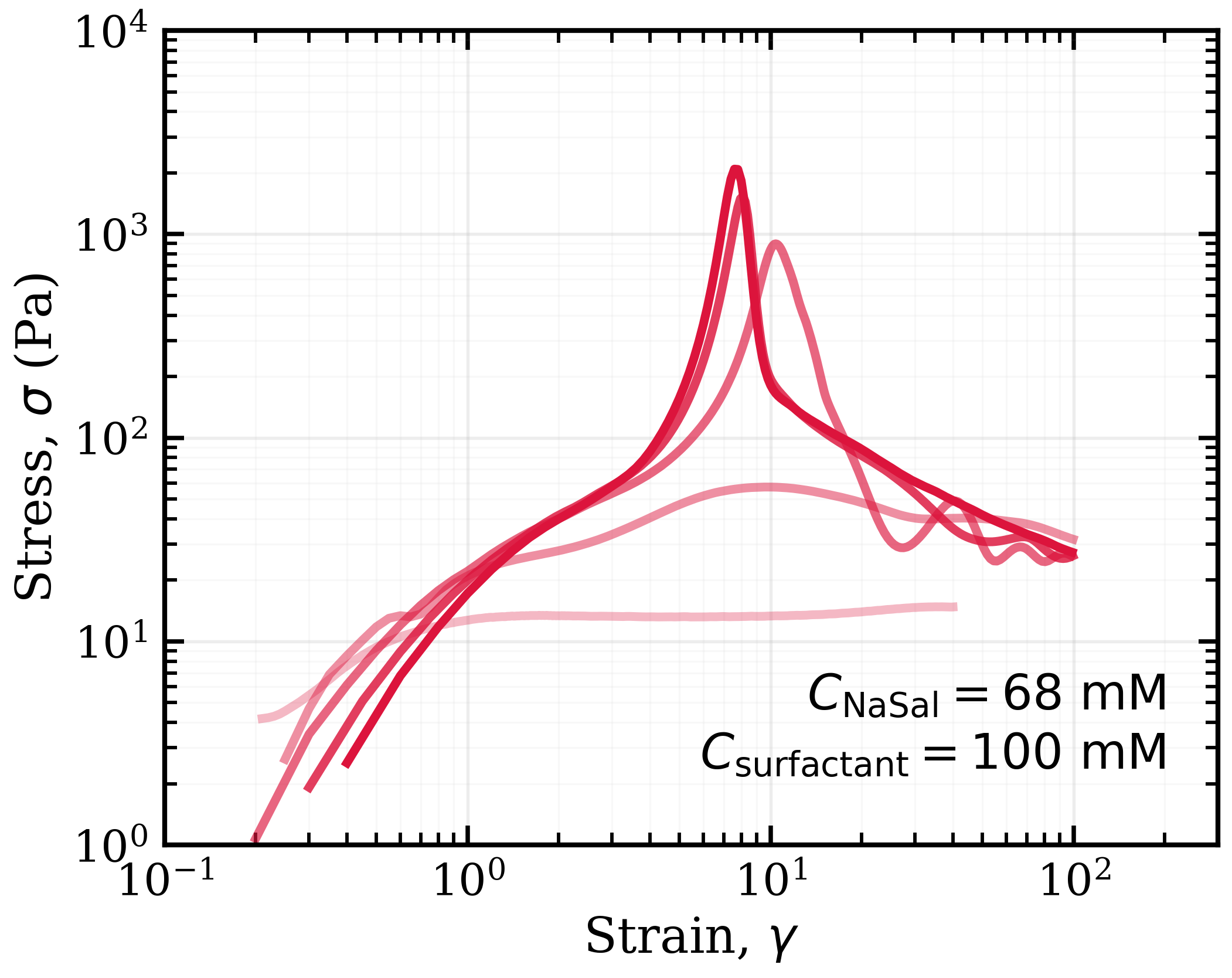}
    \end{minipage}
    \caption{Leave-one-shear-rate-out evaluation of the MFNN for the strain-hardening formulation containing $C_{\mathrm{CPyCl}}=100~\mathrm{mM}$ and $C_{\mathrm{NaSal}}=68~\mathrm{mM}$. The left panel shows the original high-fidelity (HF) experimental responses in blue, while the right panel shows the corresponding multifidelity (MF) predictions in red. Each predicted curve was generated by a separately trained model from which all HF measurements at the target shear rate had been removed. From darkest to lightest, the curves correspond to $\dot{\gamma}=20$, $15$, $10$, $5$, and $1~\mathrm{s}^{-1}$. Intermediate-rate holdouts evaluate interpolation, whereas holdouts at the limits of the retained shear-rate range evaluate extrapolation.}
    \label{fig:nasal_leave_one_gdot_out}
\end{figure}

Figure~\ref{fig:nasal_leave_one_gdot_out} presents this evaluation for the strain-hardening formulation containing $C_{\mathrm{CPyCl}}=100~\mathrm{mM}$ and $C_{\mathrm{NaSal}}=68~\mathrm{mM}$. This formulation represents a demanding test because its startup response varies strongly with the imposed shear rate. In addition to changes in the overall stress level, the experimental curves exhibit substantial variations in the development, magnitude, and position of the strain-hardening peak, followed by shear-rate-dependent post-peak relaxation.

As shown in Figure~\ref{fig:nasal_leave_one_gdot_out}, the MFNN predictions reproduce the principal characteristics of the experimental responses over the investigated shear-rate range. The model captures the initial stress growth, the development of the strain-hardening peak, the variation in peak magnitude and location with $\dot{\gamma}$, and the subsequent relaxation toward the large-strain response. The predicted curves also preserve the overall ordering and evolution of the experimental responses with increasing shear rate.

The agreement obtained for the intermediate-rate holdouts indicates that the MFNN can reconstruct complete stress-growth curves within gaps in the experimental shear-rate coverage. This result suggests that a densely sampled experimental sweep may not be necessary for every formulation. Instead, measurements at a smaller number of appropriately selected shear rates may provide sufficient high-fidelity information for the multi-fidelity model to estimate the responses at intermediate conditions. Such interpolation could reduce the experimental cost associated with mapping the full shear-rate dependence of multiple formulations.

The results for the boundary-rate holdouts provide a more rigorous test of the framework. In these cases, the model reconstructed responses at conditions lying outside the shear-rate interval retained for the target formulation. The results therefore support the use of the MFNN for limited extrapolation, particularly toward high shear rates. However, these tests should not be interpreted as evidence of unrestricted extrapolation. Prediction uncertainty is expected to increase as the target shear rate moves farther from the experimentally sampled domain, and additional high-fidelity validation remains necessary before predictions are used far beyond the investigated range.

After evaluating individual holdouts, the difficulty of the validation was increased by simultaneously removing two and then three complete shear-rate curves from the same formulation. This stage was designed to represent a more sparsely sampled experiment in which several responses are missing rather than only one. All high-fidelity points associated with the selected shear rates were removed before training, and a single MFNN was used to reconstruct all excluded curves.

Different combinations of shear rates were considered. When multiple intermediate rates were withheld, the model had to interpolate across a wider unsampled interval. These tests examined whether a small number of experimentally measured curves could support the reconstruction of several intermediate responses. Other combinations included both intermediate and boundary shear rates. In those cases, the model was required to interpolate some responses while simultaneously extrapolating others within the same training run. For example, an intermediate shear-rate curve and the highest available shear-rate curve could be removed together, producing a combined interpolation--extrapolation problem.

The multifidelity predictions continued to recover the main transient characteristics when two or three shear rates were excluded simultaneously. The predicted curves preserved the characteristic stress buildup, stress overshoot or strain-hardening peak, peak position, and post-peak relaxation observed in the corresponding experiments. As expected, the reconstruction became more challenging when the withheld conditions created a wide gap between the retained shear rates or when multiple boundary responses were excluded. Nevertheless, the overall agreement indicates that the framework can provide useful predictions even when the high-fidelity shear-rate coverage for an individual formulation is substantially reduced.

The final stage extended the holdout analysis from one formulation to the complete set of formulations within each salt family. Multiple high-fidelity curves were randomly selected from different formulation cases and removed simultaneously. For example, the excluded set could contain a high-shear-rate response from one Diclo formulation, an intermediate-shear-rate response from a second formulation, and a low-shear-rate response from a third. The model was then trained using the remaining Diclo measurements and used to reconstruct all excluded responses in a single run. The same procedure was applied independently to the NaSal family. Randomly distributing the withheld curves across multiple formulations therefore tests whether the model can handle heterogeneous gaps in the dataset rather than a single regularly structured gap.

The distributed holdouts also required interpolation and extrapolation to be performed simultaneously across different formulation cases. A curve removed from the interior of the shear-rate range of one formulation represented interpolation, whereas a boundary curve removed from another formulation represented extrapolation. The MFNN was therefore required to use information distributed across the entire salt-specific dataset to reconstruct nonlinear responses having different stress levels, peak structures, and relaxation behaviors.

\begin{figure}[t]
    \centering

    \begin{minipage}[t]{0.335\textwidth}
        \centering
        \includegraphics[
            width=\linewidth,
            trim={0 0 0 0},
            clip
        ]{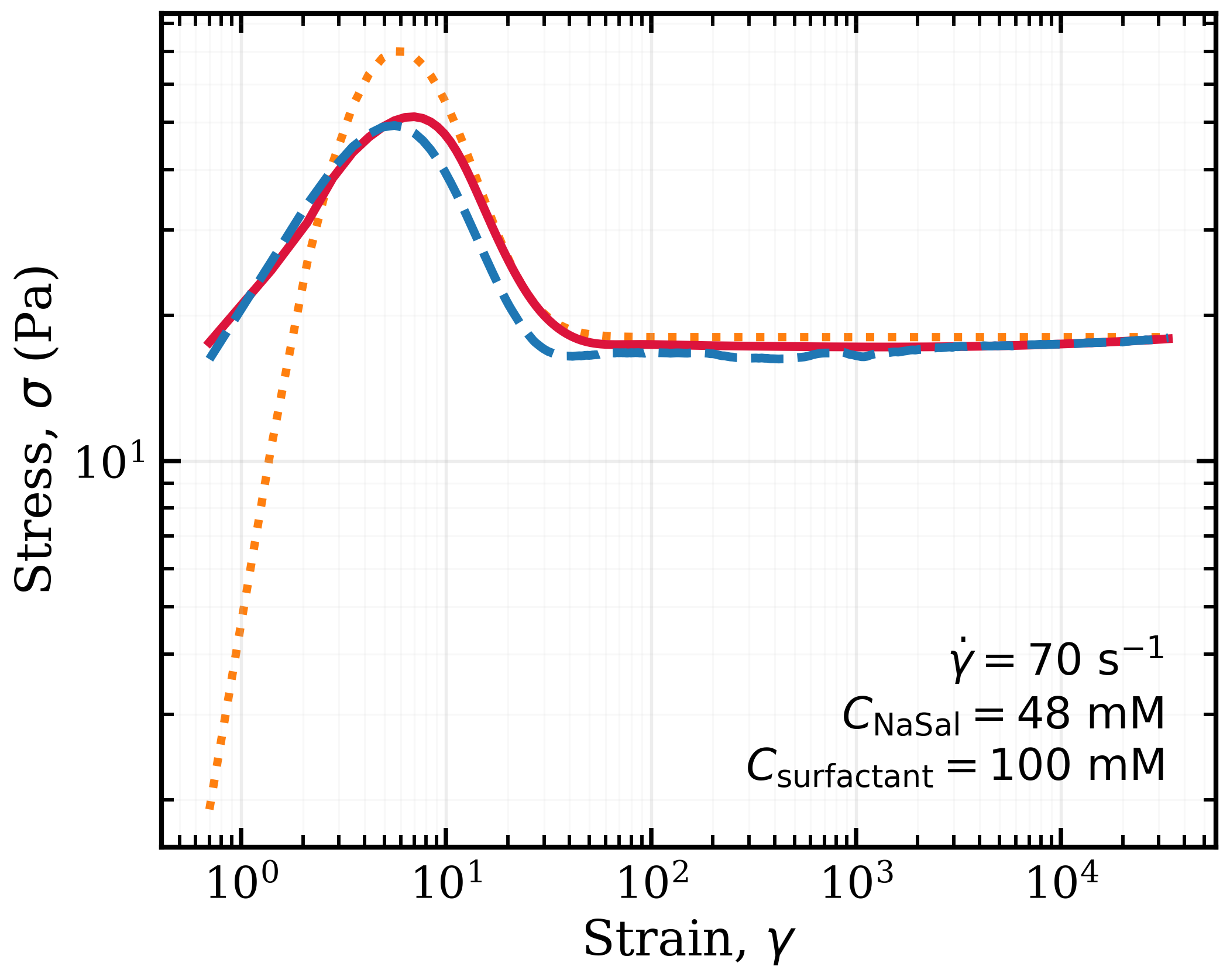}
    \end{minipage}%
    \hfill%
    \begin{minipage}[t]{0.32\textwidth}
        \centering
        \includegraphics[
            width=\linewidth,
            trim={15 0 0 0},
            clip
        ]{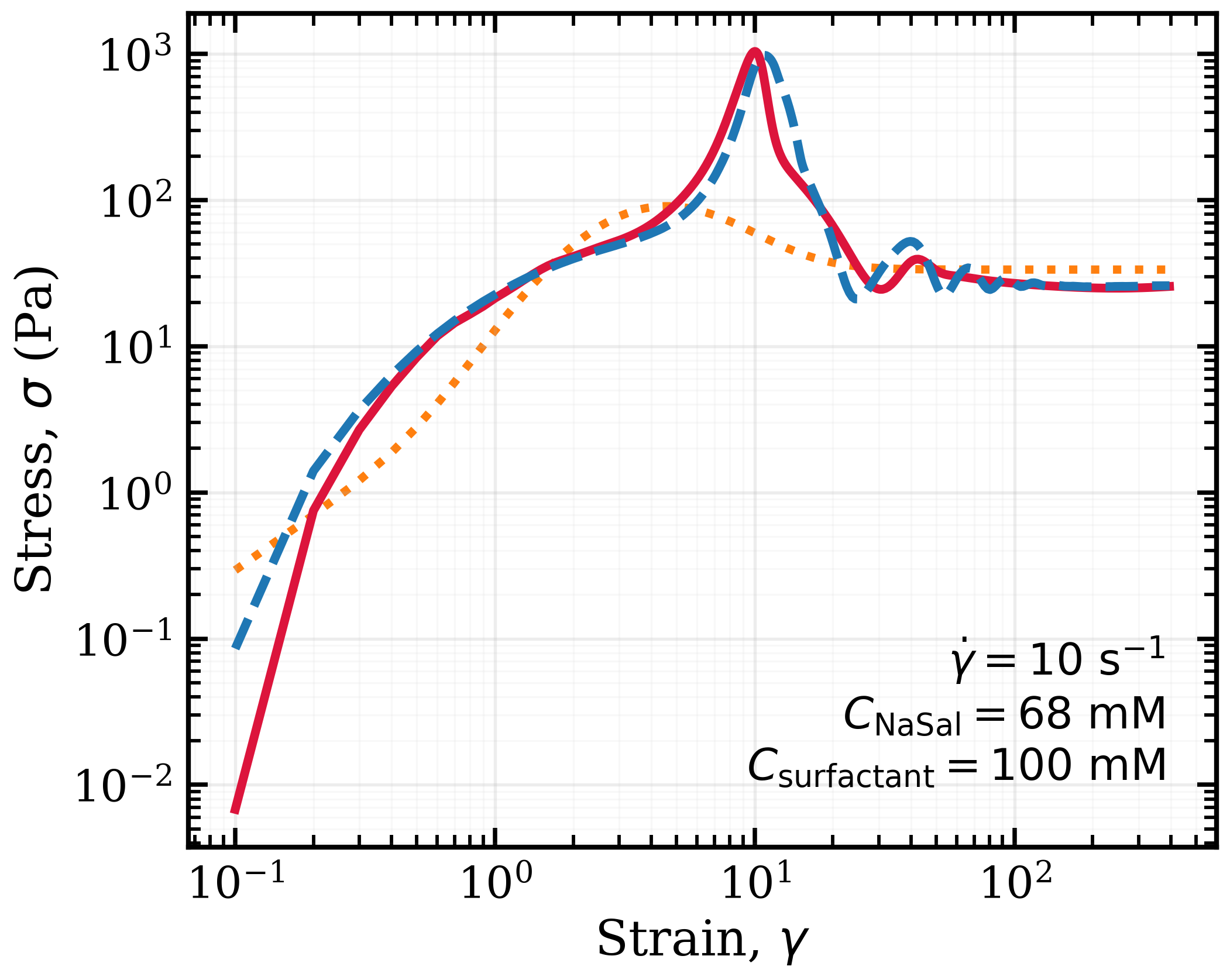}
    \end{minipage}%
    \hfill%
    \begin{minipage}[t]{0.32\textwidth}
        \centering
        \includegraphics[
            width=\linewidth,
            trim={15 0 0 0},
            clip
        ]{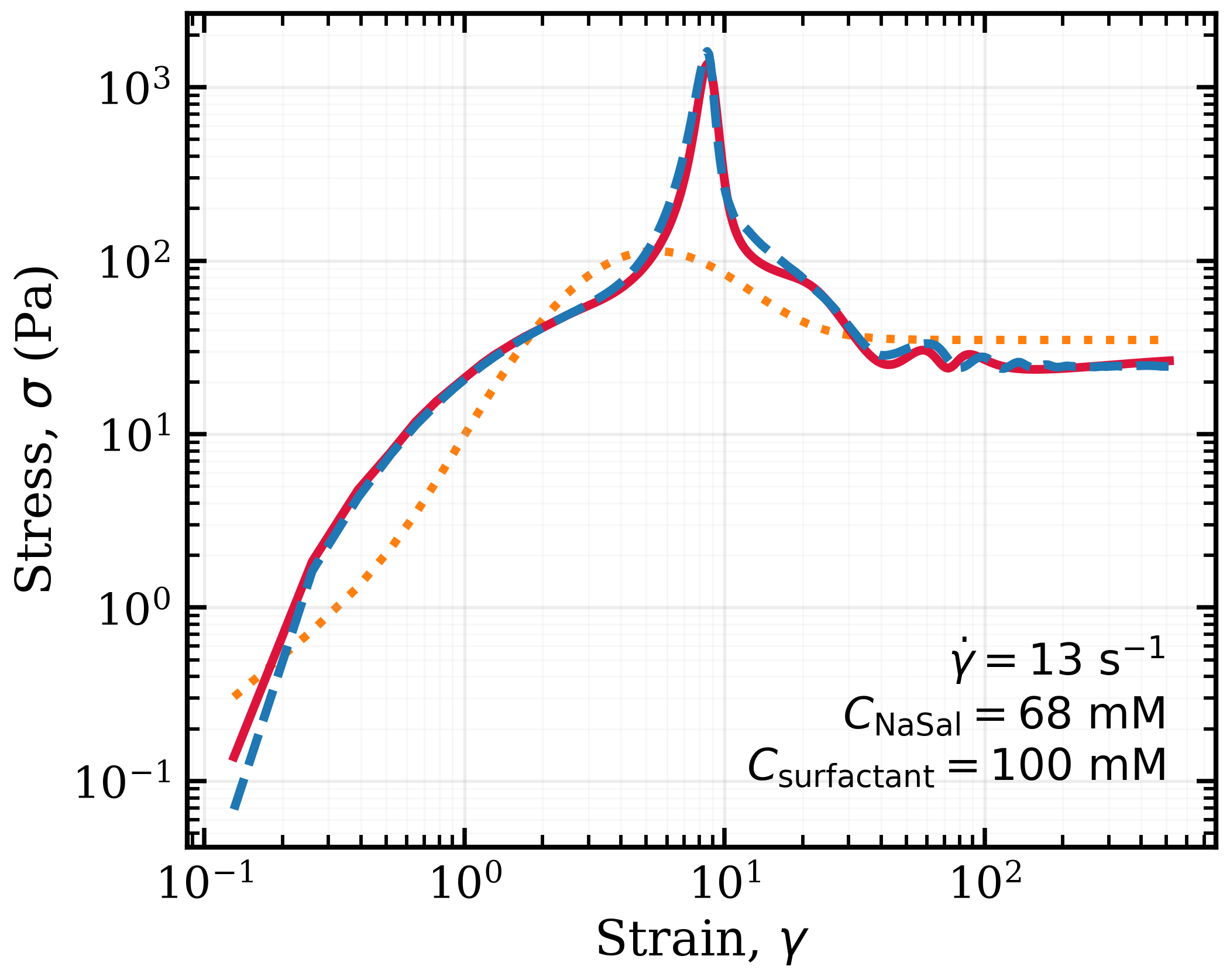}
    \end{minipage}

    \caption{Representative randomly selected leave-one-shear-rate-out evaluations for the CPyCl--NaSal system. The left panel shows a high-shear-rate extrapolation test for the non-hardening formulation containing $C_{\mathrm{CPyCl}}=100~\mathrm{mM}$ and $C_{\mathrm{NaSal}}=48~\mathrm{mM}$ at $\dot{\gamma}=70~\mathrm{s}^{-1}$. The center and right panels show interpolation tests for the strain-hardening formulation containing $C_{\mathrm{CPyCl}}=100~\mathrm{mM}$ and $C_{\mathrm{NaSal}}=68~\mathrm{mM}$ at $\dot{\gamma}=10$ and $13~\mathrm{s}^{-1}$, respectively. Each panel was generated from an independent holdout run in which the complete target high-fidelity curve was excluded from training. Blue dashed curves show the withheld high-fidelity responses, orange dotted curves show the fixed low-fidelity estimates, and red solid curves show the MFNN predictions.}
    \label{fig:random_nasal_holdouts}
\end{figure}

Figure~\ref{fig:random_nasal_holdouts} presents three representative examples from these randomly selected holdout  tests. The first two conditions correspond to the strain hardening formulation containing $C_{\mathrm{CPyCl}}=100~\mathrm{mM}$ and $C_{\mathrm{NaSal}}=68~\mathrm{mM}$ at $\dot{\gamma}=13$ and $10~\mathrm{s}^{-1}$, respectively. The third condition corresponds to the non-hardening formulation containing $C_{\mathrm{CPyCl}}=100~\mathrm{mM}$ and $C_{\mathrm{NaSal}}=48~\mathrm{mM}$ at $\dot{\gamma}=70~\mathrm{s}^{-1}$.

The predictions obtained from these multicase holdouts remained in good agreement with the corresponding withheld experimental curves. The model reconstructed both comparatively smooth non-hardening responses and strongly nonlinear strain-hardening curves containing localized stress peaks. This result indicates that the multifidelity framework can make use of shared trends across the formulation space while retaining the formulation-specific corrections required to reproduce different nonlinear response types.

It is important to distinguish this distributed shear-rate holdout from the prediction of a completely unseen formulation. In the present test, individual shear-rate curves were removed from several formulations, but other high-fidelity curves from those formulations generally remained in the training dataset. The test therefore evaluates robustness to missing shear-rate measurements distributed across the formulation space. Removing all shear-rate curves associated with one CPyCl--salt concentration pair would instead constitute a formulation-level holdout and a separate evaluation of generalization to an unseen composition.

Taken together, the three stages provide a progressive assessment of the multifidelity framework under increasingly severe forms of experimental sparsity. The single-curve holdouts evaluate isolated interpolation and extrapolation. The simultaneous two- and three-curve holdouts determine whether multiple responses can be reconstructed from a reduced set of experiments for one formulation. The randomly selected holdout tests broaden the evaluation across different formulations, shear rates, and response regimes within each salt family.

These capabilities are particularly relevant for extracting nonlinear rheological quantities such as peak stress, peak strain, hardening-onset strain, and the shear-rate interval over which strain hardening occurs. Reliable extraction of these quantities requires the full transient stress-growth curve rather than isolated stress measurements. By reconstructing complete responses at unmeasured shear rates, the MFNN can augment sparsely sampled experimental datasets, reduce the need for exhaustive shear-rate sweeps, and help identify regions where additional high-fidelity measurements would be most informative.

The predicted curves should nevertheless be regarded as experimentally informed surrogate responses rather than unconditional replacements for experimental measurements. Interpolation is supported by high-fidelity data on both sides of the target condition and is therefore generally expected to be more reliable. Extrapolation, particularly far beyond the measured shear-rate range, carries greater uncertainty and should be accompanied by targeted experimental validation whenever possible. Within these limitations, the results demonstrate that the multifidelity framework provides a practical strategy for extending incomplete rheological datasets across both experimentally expensive intermediate conditions and difficult-to-access high-shear-rate regimes.

Following the shear-rate holdout evaluation, the generalization capability of the multifidelity framework was examined using a more demanding formulation-level holdout test. The purpose of this analysis was to determine whether the model could predict the complete startup behavior of a formulation for which no high-fidelity measurements were available at any imposed shear rate. This test represents the practical situation in which a particular combination of surfactant and salt concentrations has not been characterized experimentally, but its nonlinear rheological response is required over a range of shear rates.

For each formulation-level holdout, one complete combination of CPyCl and salt concentrations was selected, and all high-fidelity startup curves associated with that formulation were removed before training. Thus, unlike the shear-rate holdout analysis, the model did not retain any experimental response from the target formulation. A single MFNN was trained using the remaining formulations within the corresponding salt family and was subsequently evaluated at the excluded concentration pair over multiple imposed shear rates. The withheld experimental curves were used only as reference data for evaluating the predictions.

The low-fidelity responses at the target conditions provided a structured representation of the general startup behavior, while the multifidelity correction was inferred from the remaining high-fidelity formulation data. This validation therefore examines whether information learned from neighboring compositions can be transferred to a completely excluded formulation. It also provides a substantially more stringent test than withholding individual shear rates because the model cannot rely on any high-fidelity curve from the target concentration pair.

In addition to reconstructing the stress magnitude, the model must recover the appropriate nonlinear response regime of the excluded formulation. In particular, it must determine whether the startup response remains comparatively smooth and non-hardening or develops a localized strain-hardening peak followed by post-peak relaxation. Hardening and non-hardening labels are not supplied to the network as explicit targets. Instead, the response regime emerges from the predicted stress-growth curve and can subsequently be identified from features such as the magnitude and location of the stress peak, the rapidity of stress amplification, and the post-peak evolution.

Depending on the position of the excluded concentrations relative to the retained formulations, this test can represent interpolation or extrapolation in composition space. When the target concentrations lie between experimentally characterized formulations, the model performs compositional interpolation. When the target lies outside the retained concentration range, prediction requires compositional extrapolation. In both cases, the model must simultaneously reconstruct the shear-rate dependence of the complete transient response without access to high-fidelity measurements from the target formulation.

\begin{figure}[t]
    \centering
    \begin{minipage}[t]{0.49\textwidth}
        \centering
        \includegraphics[
            width=\linewidth,
            trim={0 15 0 0},
            clip
        ]{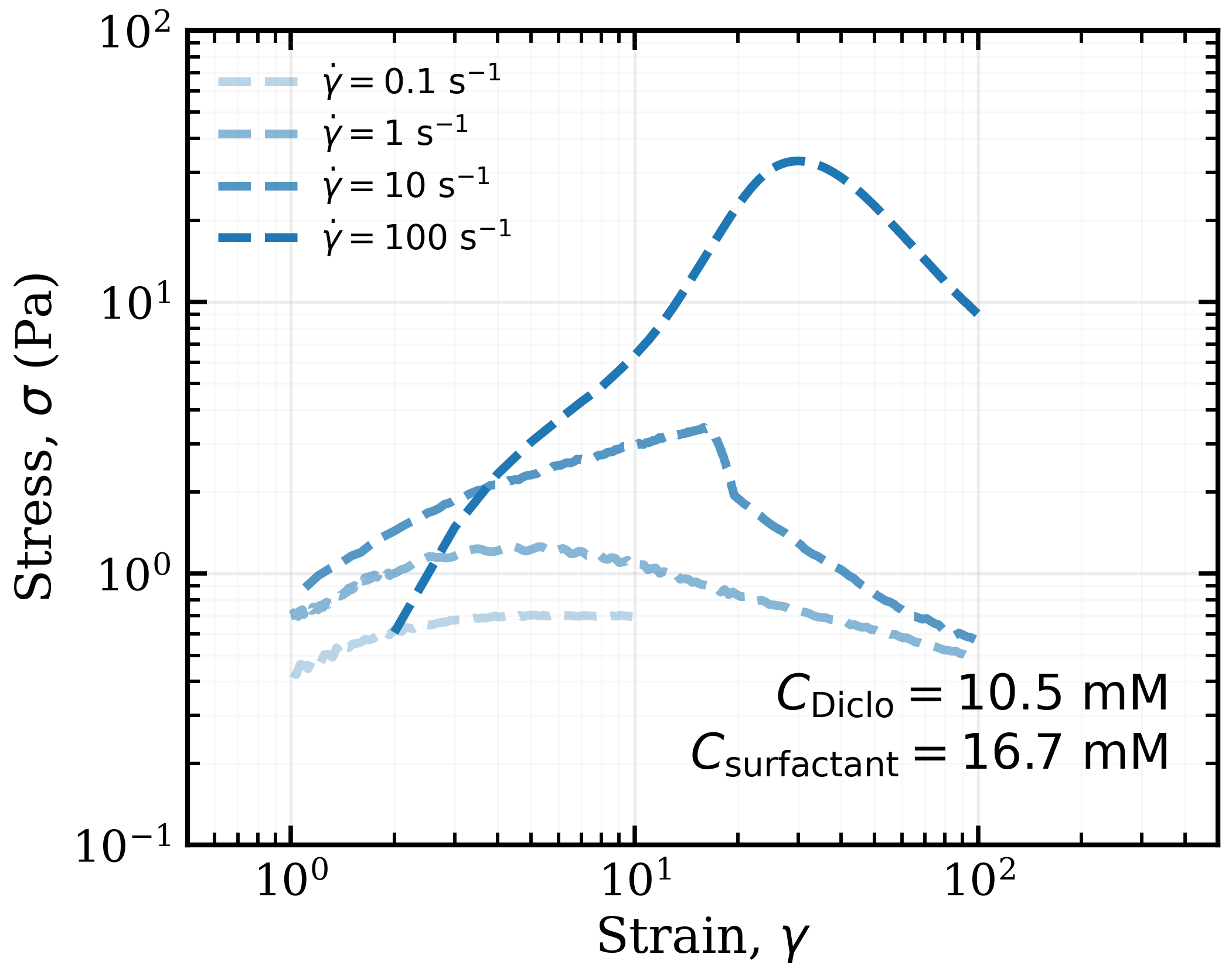}
    \end{minipage}%
    \hfill%
    \begin{minipage}[t]{0.50\textwidth}
        \centering
        \includegraphics[
            width=\linewidth,
            trim={0 15 0 0},
            clip
        ]{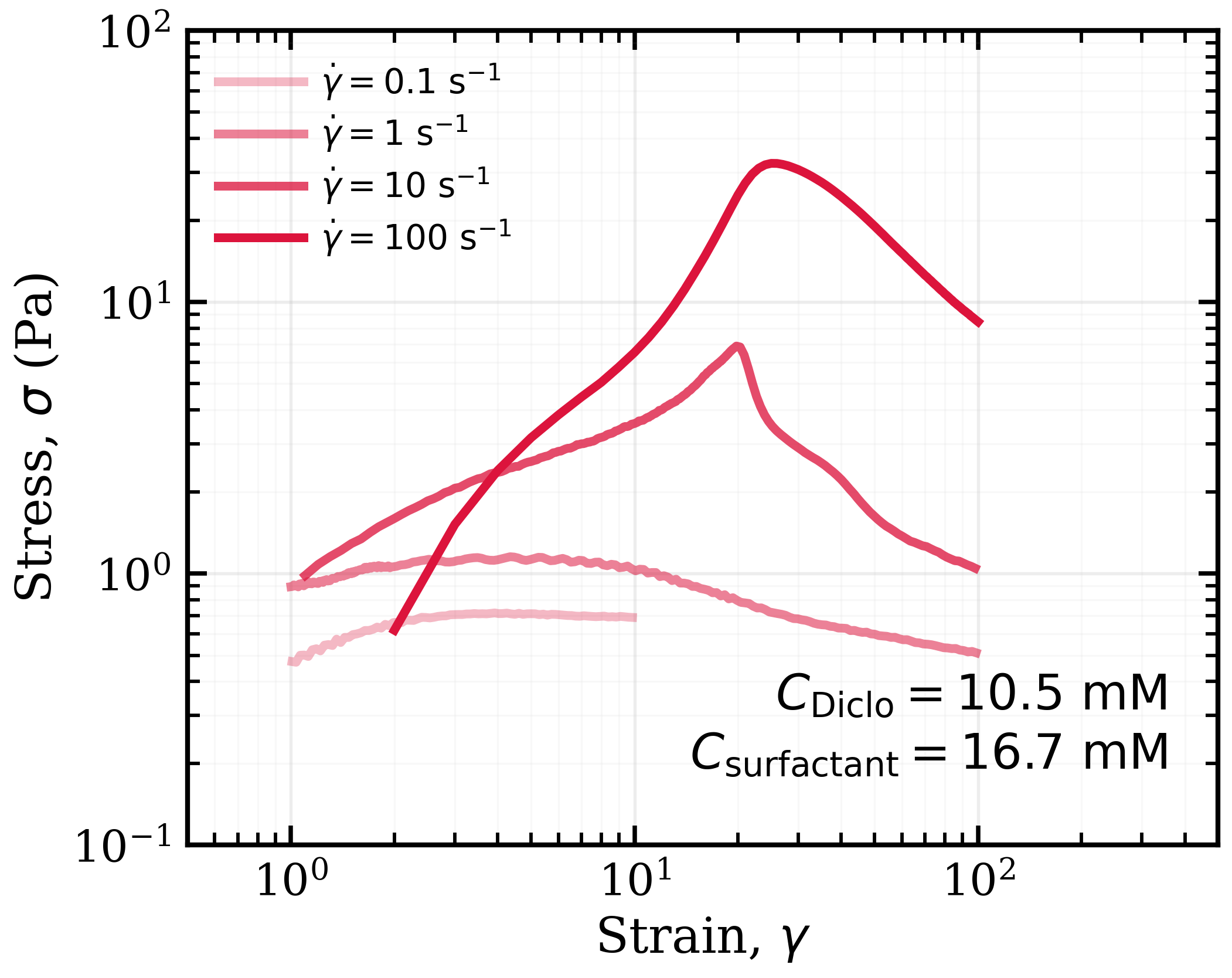}
    \end{minipage}
    \begin{minipage}[t]{0.49\textwidth}
        \centering
        \includegraphics[
            width=\linewidth,
            trim={0 0 0 0},
            clip
        ]{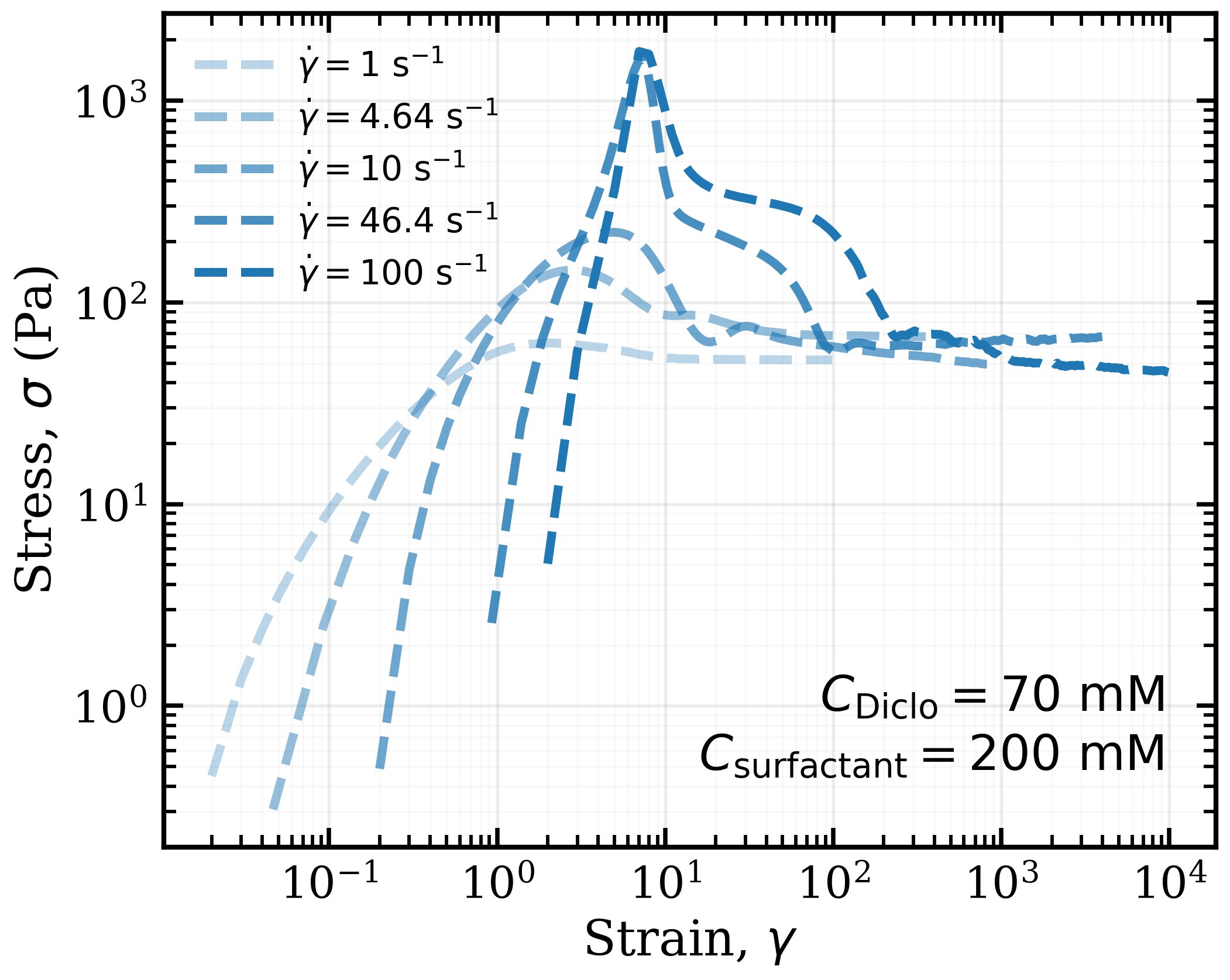}
    \end{minipage}%
    \hfill%
    \begin{minipage}[t]{0.49\textwidth}
        \centering
        \includegraphics[
            width=\linewidth,
            trim={0 0 0 0},
            clip
        ]{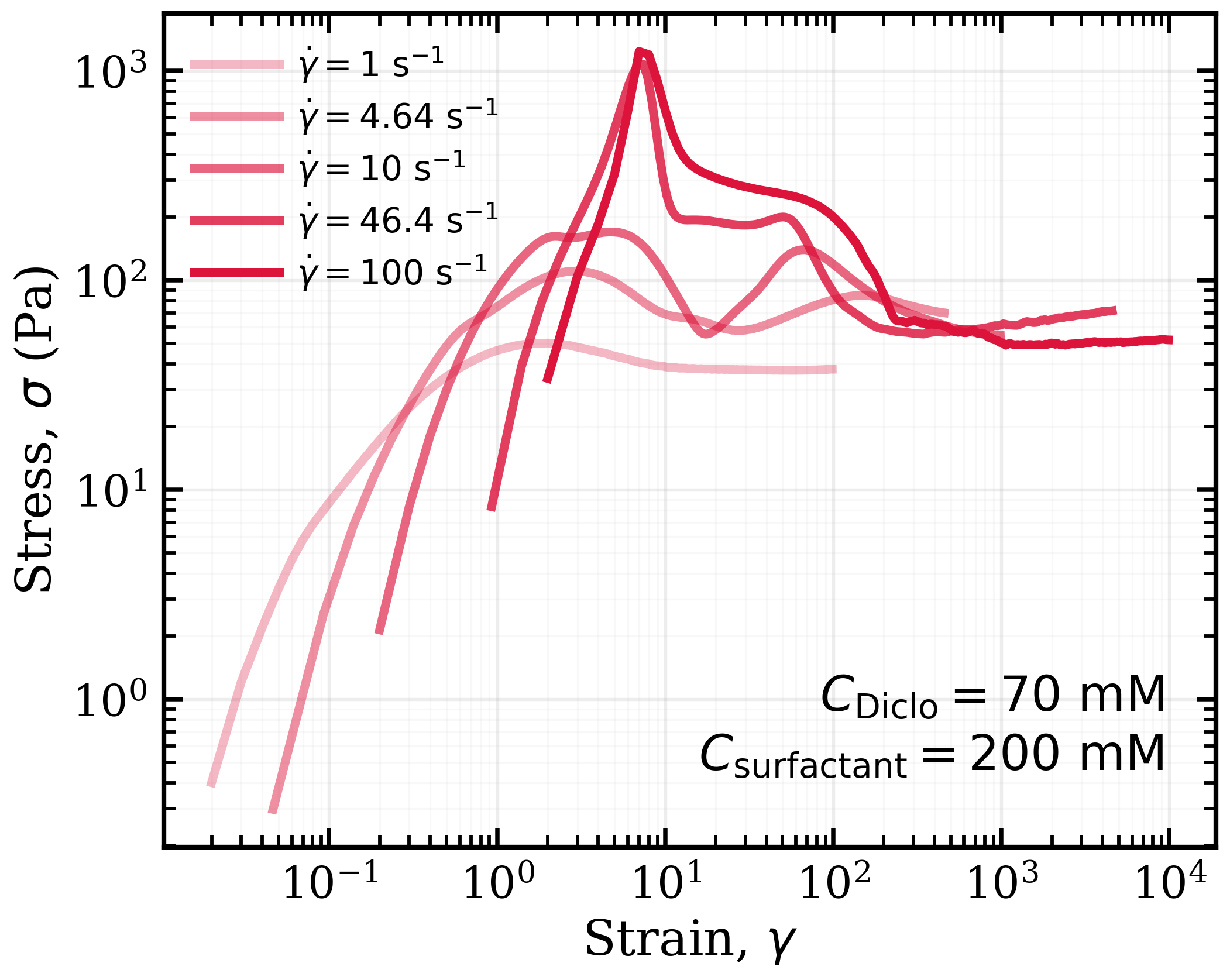}
    \end{minipage}
    \caption{Formulation-level holdout evaluation for two CPyCl--Diclo systems. The upper row presents the formulation containing $C_{\mathrm{CPyCl}}=16.7~\mathrm{mM}$ and $C_{\mathrm{Diclo}}=10.5~\mathrm{mM}$, while the lower row presents the formulation containing $C_{\mathrm{CPyCl}}=200~\mathrm{mM}$ and $C_{\mathrm{Diclo}}=70~\mathrm{mM}$. In each row, the left panel shows the withheld high-fidelity experimental responses and the right panel shows the corresponding displayed multifidelity responses.}
    \label{fig:heldout_diclo_formulations}
\end{figure}

Figure~\ref{fig:heldout_diclo_formulations} presents two representative formulation-level holdout tests for the CPyCl--Diclo system. In each row, the left panel shows the withheld high-fidelity experimental responses and the right panel shows the corresponding MFNN predictions. All curves associated with each target concentration pair were excluded from training, and the multiple shear-rate predictions in each row were generated using a single model trained without that formulation.

The first holdout case contains $C_{\mathrm{CPyCl}}=16.7~\mathrm{mM}$ and $C_{\mathrm{Diclo}}=10.5~\mathrm{mM}$. For the same CPyCl concentration, the available neighboring formulations at $C_{\mathrm{Diclo}}=10$ and $11~\mathrm{mM}$ both exhibit strain-hardening behavior. The excluded $10.5~\mathrm{mM}$ condition therefore represents interpolation within a locally consistent hardening region of composition space. As shown in the upper row of Figure~\ref{fig:heldout_diclo_formulations}, the MFNN predicts the emergence of pronounced stress peaks and their variation with shear rate. The model also recovers the main features of the post-peak response, demonstrating that it can transfer the nonlinear trends learned from the neighboring hardening formulations to an unmeasured intermediate concentration.

The second holdout formulation, containing $C_{\mathrm{CPyCl}}=200~\mathrm{mM}$ and $C_{\mathrm{Diclo}}=70~\mathrm{mM}$, provides a more challenging and informative test. At this CPyCl concentration, the dependence of the startup response on Diclo concentration is nonmonotonic. The formulation containing $60~\mathrm{mM}$ Diclo does not exhibit strain hardening, whereas the formulations containing $75$ and $80~\mathrm{mM}$ Diclo display pronounced hardening. The response at $90~\mathrm{mM}$ again belongs to the non-hardening category. Consequently, the model cannot determine the behavior at $70~\mathrm{mM}$ by assuming a simple monotonic transition with increasing salt concentration.

Despite the complete removal of the $200~\mathrm{mM}$ CPyCl--$70~\mathrm{mM}$ Diclo formulation, the MFNN reconstructs the formation of the experimentally observed strain-hardening peaks over the relevant shear-rate range. This result is particularly significant because the target formulation lies near a change in nonlinear response regime. The prediction indicates that the model has learned composition-dependent relationships that extend beyond a direct interpolation of stress magnitude and can recover whether localized hardening should emerge at the excluded concentration.

The predicted and experimental curves do not coincide perfectly at every strain and shear rate. Some deviations remain in the precise peak magnitude, peak width, post-peak relaxation, and large-strain stress level. Nevertheless, the model recovers the features most relevant to identifying and characterizing strain hardening. 

This feature-level agreement is important because the principal objective is not limited to pointwise reconstruction of stress. The predicted curves are subsequently used to extract nonlinear quantities such as peak stress, peak strain, hardening-onset strain, and the shear-rate interval over which hardening occurs. Recovering the correct response category and the approximate location of the associated transient features therefore provides physically meaningful information even when small pointwise discrepancies remain.

The same formulation-level holdout procedure was applied to additional CPyCl--Diclo and CPyCl--NaSal formulations exhibiting both hardening and non-hardening responses. These tests assess whether the framework can reconstruct an unmeasured composition and determine its nonlinear response regime from the remaining formulation data. The results extend the earlier shear-rate holdout analysis: whereas the shear-rate tests retained other high-fidelity curves from the target formulation, the formulation-level tests require information to be transferred entirely from other compositions within the same salt family.

From an experimental perspective, this capability may reduce the number of formulations that require complete characterization. Measurements could be concentrated on a selected set of representative compositions, while the validated multifidelity model is used to estimate responses at intermediate concentration pairs and identify regions where the nonlinear behavior changes. These predictions could then guide targeted experiments near transitions between non-hardening and strain-hardening regimes. Predictions involving concentration extrapolation or conditions far from the available high-fidelity formulation space should, however, be interpreted cautiously and validated experimentally whenever possible. Beyond its practical value, a systematic investigation of concentration effects could also help reveal physical connections that are still not fully understood. In particular, it may help clarify how nonlinear rheological responses scale with the number of entanglements, how changes in microstructure, including transitions between predominantly linear and branched micellar networks, influence the onset of strain hardening, and what role micellar persistence length plays in governing the mechanical response. Such an approach could also provide insight into the conditions under which the elastic energy accumulated during shear deformation becomes comparable to the energy required for micellar scission, potentially triggering flow-induced micelle breakage and transitions in the observed nonlinear behavior. Ultimately, these relationships could contribute to a more physically grounded interpretation of the concentration-dependent response captured by the model

\section{Conclusions}

In this work, a multi-fidelity machine-learning framework was developed to predict complete startup stress-growth responses of CPyCl-based wormlike micellar solutions from limited and nonuniformly distributed rheological measurements. The framework combines a fixed analytical representation, which supplies the dominant structure of the startup response, with high-fidelity experimental data used to learn formulation- and shear-rate-dependent nonlinear corrections. By formulating the learning problem in logarithmic stress space, the model accommodates the broad range of stress magnitudes encountered across two different salt (NaSal and Diclo) systems.

The predictive capability of the framework was evaluated using progressively more demanding holdout tests. In the shear-rate holdout analysis, complete experimental curves were excluded rather than individual stress points. The model reconstructed intermediate responses through interpolation and provided reasonable predictions at the boundaries of the retained shear-rate range. The latter tests represent controlled assessments of limited extrapolation and are particularly relevant at high shear rates, where startup experiments may be difficult because of short transient times, limited temporal resolution, instrument inertia, or flow instabilities. Simultaneous removal of multiple shear-rate curves further showed that the framework can recover several missing responses when experimental coverage within an individual formulation is substantially reduced.

The distributed multicase holdout tests extended this evaluation to heterogeneous gaps across each salt family. Under these conditions, multiple shear-rate curves were removed from different formulations, requiring interpolation and extrapolation to be performed simultaneously across the dataset. The reconstructed responses retained the principal transient features observed experimentally, including initial stress buildup, stress overshoot, localized strain-hardening peaks, and post-peak relaxation. These results demonstrate that the model can use information distributed across the formulation space while preserving important formulation-specific nonlinear behavior.

A more stringent evaluation was performed by withholding complete formulations, such that no high-fidelity startup curve from the target concentration pair was available during training. For the representative CPyCl--Diclo cases examined, the model recovered the qualitative shear-rate dependence and the experimentally observed strain-hardening regime of the excluded formulations. Of particular interest was the prediction for the formulation containing $200\,\mathrm{mM}$ CPyCl and $70\,\mathrm{mM}$ Diclo, which lies near a transition between non-hardening and hardening responses as salt concentration is varied. Although pointwise deviations remained in quantities such as the exact peak magnitude, peak width, and large-strain stress level, the predicted curves recovered whether hardening occurred and approximately where its principal transient features appeared.

These findings indicate that the proposed framework can provide an experimentally informed surrogate for augmenting sparse startup-rheology datasets. Once validated within the relevant material and operating domain, the model may reduce the need for densely sampled shear-rate sweeps, provide estimates at experimentally expensive intermediate conditions, and guide measurements in difficult-to-access high-shear-rate regimes. Prediction of complete stress-growth curves also provides a basis for extracting nonlinear quantities such as peak stress, peak strain, hardening-onset strain, and the shear-rate interval over which hardening occurs.

It should also be noted that the present conclusions remain limited by the size and nonuniform coverage of the available experimental dataset. Interpolation is supported by neighboring high-fidelity measurements and is generally more reliable than extrapolation, whereas predictions far beyond the measured shear-rate or concentration ranges require additional experimental validation. Future work should expand the dataset across salt chemistry, concentration, and shear rate; introduce uncertainty quantification; and evaluate the framework against independently acquired formulations. The resulting augmented response maps may then be used to investigate quantitative relationships among formulation variables, linear viscoelastic properties, and nonlinear strain-hardening metrics, ultimately supporting the data-efficient design of wormlike micellar systems with targeted transient rheological behavior. They may also help uncover the physics underlying the nonlinear response, including the roles of entanglement density, micellar persistence length, and transitions between linear and branched micellar microstructures, as well as the onset of flow-induced micellar scission. It should also be clearly noted that multi-fidelity data-driven approaches, while powerful in fusing different modes/fidelities of data, are ultimately unable to predict behavior in other untested flow protocols or conditions as no explicit physical intuition is being embedded. 

\section*{Acknowledgements}
SZ and SJ would like to thank the Office of Naval Research (ONR) for funding from MURI \#N00014-23-1-2499.


\printbibliography

\end{document}